\documentclass[10pt,journal,a4paper,twoside,twocolumn]{IEEEtran}
\usepackage{cite}
\usepackage{float}
\usepackage{stfloats}
\usepackage{fancyhdr}
\usepackage{color}
\usepackage{makecell}
\usepackage[inline]{enumitem}
\usepackage{amsmath,amsfonts,amssymb}
\usepackage{graphicx}
\usepackage{algorithm}
\usepackage{algorithmic}
\usepackage{epsfig}
\usepackage{subfigure}
\usepackage{epstopdf}
\usepackage{bm}
\usepackage{array}
\usepackage{bbding}
\usepackage{booktabs}
\usepackage{multirow}
\usepackage{threeparttable}

\usepackage{bbm}

\newtheorem{Lemma}{Lemma}
\newtheorem{Rem}{Remark}

\begin{document}

\title{
	Low-Complexity Null-Space-Based Simultaneous Wireless Information and Power Transfer Scheme
}

\author{\IEEEauthorblockN{
	Cheng Luo, \IEEEmembership{Graduate Student Member, IEEE}, Jie Hu, \IEEEmembership{Senior Member, IEEE}, Luping Xiang, \IEEEmembership{Senior Member, IEEE}, Kun Yang, \IEEEmembership{Fellow, IEEE} and Zhiqin Wang}
	\\

	\thanks{Cheng Luo and Jie Hu are with the School of Information and Communication Engineering, University of Electronic Science and Technology of China, Chengdu, 611731, China, email: chengluo@std.uestc.edu.cn; hujie@uestc.edu.cn}
	\thanks{Luping Xiang is with the State Key Laboratory of Novel Software Technology, Nanjing University, Nanjing 210008, China, and also with the School of Intelligent Software and Engineering, Nanjing University (Suzhou Campus), Suzhou 215163, China, e-mail: luping.xiang@nju.edu.cn}
	\thanks{Kun Yang is with the School of Computer Science and Electronic Engineering, University of Essex, Colchester CO4 3SQ, U.K., email: kunyang@essex.ac.uk}
	\thanks{Zhiqin Wang is with the China Academy of information and Communications Technology, Beijing, 100191, China, e-mail: wangzhiqin@caict.ac.cn}
}

\maketitle

\thispagestyle{fancy} 
\lhead{} 
\chead{} 
\rhead{} 
\lfoot{} 
\cfoot{} 
\rfoot{\thepage} 
\renewcommand{\headrulewidth}{0pt} 
\renewcommand{\footrulewidth}{0pt} 
\pagestyle{fancy}

\rfoot{\thepage} %

\begin{abstract}
	Simultaneous wireless information and power transfer (SWIPT) has attracted sustained interest. We propose a null-space-based transmission scheme for multiuser SWIPT serving both energy users (EUs) and information users (IUs). Under a practical nonlinear energy-harvesting (EH) model and multiple waveform options, we revisit the role of dedicated energy beams (EBs). We show that, in general, dedicated EBs are unnecessary because information beams (IBs) with Gaussian signaling can simultaneously support wireless energy transfer (WET) and wireless information transfer (WIT), unless special energy-centric waveforms (e.g., deterministic sinusoidal waveforms) are employed and provide sufficient gains. Guided by these insights, we formulate an optimization problem for EB design to enable dedicated waveform transmission for WET, and we develop a low-complexity algorithm that reduces computation by ignoring the WET contribution of IBs during optimization. Numerical results corroborate that deterministic sinusoidal waveforms outperform Gaussian signaling when the received RF power lies in the EH high-efficiency region, making dedicated EBs beneficial. The proposed scheme achieves computational complexity reductions of 91.43\% and 98.54\% for the cases $M=8,,K^I=K^E=2$ and $M=16,,K^I=K^E=4$, respectively, with negligible performance loss, thereby validating the efficiency of the low-complexity algorithm.
\end{abstract}

\begin{IEEEkeywords}
	Simultaneous wireless information and power transfer, null space, practical energy receiver, low complexity design.
\end{IEEEkeywords}
\section{Introduction}

\IEEEPARstart{I}{n} the forthcoming sixth-generation (6G) communication networks, a vast array of internet of things (IoT) devices and sensors will be deployed across a wide spectrum of applications, including healthcare, environmental monitoring, smart homes, smart cities, autonomous transportation, national defense, and other demanding environments\cite{Bruno_intro, Bruno_2}. Studies suggest that IoT device density within these future networks may reach tens or more of devices per square meter\cite{Onel_intro}, necessitating both robust connectivity and highly energy-efficient operation. Guided by Koomey's law, these devices are anticipated to complete tasks with significantly reduced energy consumption\cite{Koomey_intro}, which further drives the realization of future IoT visions. However, this surge in low-power, high-maintenance devices presents challenges for next-generation wireless network design and emphasizes the importance of exploring advanced solutions, such as wireless energy transfer (WET)\cite{TWC_WET1, TWC_WET2}, simultaneous wireless information and power transfer (SWIPT)\cite{TWC_SWIPT1, TWC_SWIPT2} and wirelessly powered communication networks (WPCNs)\cite{TWC_WPCNs, RISS_luo}, to sustainably and efficiently address these demands. Moreover, emerging application domains, such as human digital twin–enabled personalized healthcare, vehicular networks, and energy-sustainable IoT deployments in smart cities further underscore the need for efficient co-transfer of energy and information, for which SWIPT-based architectures offer a compelling pathway to long-lived, always-connected devices\cite{chen2023networking, chen2024revolution, chen2024generative}.

Extensive research on WET and SWIPT have explored various domains, including modulation and coding techniques\cite{Cuijinwen, Hujie_intro}, waveform optimization \cite{Buro_sine, waveform_YQD}, beamforming strategies\cite{Wqq_WET, JieXu_NoEB}, and novel architectures\cite{Xiaolin, wqq_RIS, pan_RIS, massive_luo, multi_RISS, xiang2024endtoend}. Specifically, \cite{Cuijinwen} introduces an end-to-end polar coding design for SWIPT systems. This approach replaces traditional encoding, decoding, modulation, demodulation, and energy harvesting (EH) components with neural networks trained as an autoencoder, significantly enhancing performance in both wireless information transfer (WIT) and WET. Moreover, a trade-off between WIT and WET in terms of coding and modulation is investigated in \cite{Hujie_intro}, providing valuable design insights for SWIPT systems. In addition, waveform selection has proven crucial for WET optimization. \cite{waveform_YQD} examines the use of M-QAM, Gaussian, and deterministic sinusoidal waveforms, each with distinct advantages and drawbacks due to the non-linear characteristics of EH circuits. Similarly, \cite{Buro_sine} compares single-tone and multi-tone sine waveforms, emphasizing the benefits of carefully designed rectifiers and waveforms. Information beamforming (IB) also emerges as a dual-function enabler for both WIT and WET, as highlighted in \cite{JieXu_NoEB} and \cite{Wqq_WET}. The results indicate that while energy beamforming (EB) may require significant resources for interference suppression, IB can serve both functions more efficiently. However, in scenarios where interference can be managed at the receiver, dedicated EB yields substantial WET benefits. Moreover, innovative architectures such as fluid antennas\cite{Xiaolin} and reconfigurable intelligent surfaces (RIS) have demonstrated outstanding SWIPT performance improvements \cite{wqq_RIS, pan_RIS, massive_luo, multi_RISS}. These advancements mark critical steps in addressing the challenges associated with integrating WIT and WET in modern communication networks.

The null space, representing the vector set yielding a zero vector upon matrix transformation, has been extensively studied and applied in diverse fields. For instance, the multiple signal classification (MUSIC) algorithm \cite{music}, a high-resolution subspace method for frequency estimation and radio direction finding, leverages the orthogonality between noise and signal subspaces. This makes it well-suited for both research and practical hardware applications. In \cite{interfere_relays}, the null space effectively suppresses self-interference in full-duplex relays, enhancing achievable rate performance. A notable joint beamforming and broadcasting scheme for massive multiple-input multiple-output (MIMO) systems is proposed in \cite{JBB}, wherein the base station (BS) positions broadcast signals in the null space of beamformed signals for information users (IUs) without channel state information (CSI), thereby preventing interference. Moreover, an energy efficiency optimization framework for coordinated multi-point SWIPT systems is proposed in \cite{CoMP-SWIPT}, which employs the null space to suppress interference across both macro and small cells, achieving a balance between transmission rate and energy harvesting requirements. Partial related works are presented in Table \ref{table:relatedworks11}.
\begin{table*}[]
    \centering
	\caption{Relevant works on SWIPT and null-space-based interference.}
	\begin{tabular}{m{3 cm}<{\centering}|c|c|c|c|c}
    \hline
    \textbf{} &  \textbf{Type}&\textbf{Interference suppression}& \textbf{Practical EH} &\textbf{Different Waveforms} & \textbf{Low complexity design} \\
    \hline
    Our Proposed&SWIPT&null-space-based&\CheckmarkBold&\CheckmarkBold&\CheckmarkBold\\
	\hline
	Self-interference\cite{interfere_relays}&WIT&null-space-based&\XSolidBrush&\XSolidBrush&\XSolidBrush\\
	\hline
	Joint beamforming and broadcasting\cite{JBB}&WIT &null-space-based& \XSolidBrush&\XSolidBrush&\XSolidBrush\\
    \hline
	Xu, et al. \cite{JieXu_NoEB}&SWIPT&optimization problem/cost-free&\XSolidBrush&\XSolidBrush&\XSolidBrush\\
	\hline
	Wu, et al.  \cite{Wqq_WET}&SWIPT&optimization problem&\XSolidBrush&\XSolidBrush&\XSolidBrush\\
	\hline
	Coordinated multi-point SWIPT\cite{CoMP-SWIPT}&SWIPT &null-space-based& \XSolidBrush&\XSolidBrush&\XSolidBrush\\
	\hline
	Zhao, et al.\cite{single_IU}&SWIPT &\XSolidBrush, single user& \CheckmarkBold&\CheckmarkBold&\CheckmarkBold\\
	\hline
    \end{tabular} \label{table:relatedworks11}
\end{table*}

\begin{figure}
	\centering
	\includegraphics[width=0.9\linewidth]{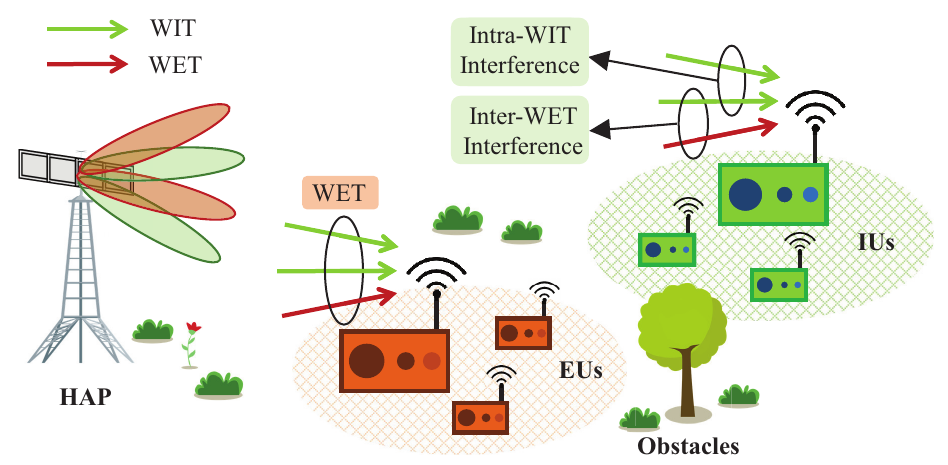}
	\caption{System model of the proposed null-space-based SWIPT system, where WET benefits from both EBs and IBs, and IUs encounter interference from intra-WIT and inter-WET sources.}
	\label{fig:systemmodel}
\end{figure}

As a step ahead from existing works, we propose a null-space-based SWIPT framework that reassess the necessity of dedicated EBs and while integrating practical EH models and waveform designs. Moreover, we design a low complexity algorithm for the corresponding problem. Our contributions can be summarized as follows:
\begin{itemize}
	\item We consider a null-space-based SWIPT system and propose a detailed interference suppression scheme to manage intra-WIT and inter-WET interferences. Our approach incorporates practical EH models and waveform designs, along with optimized power allocation schemes and a low computational complexity algorithm to enhance overall system performance.

	\item We reassess the necessity of dedicated EBs in the proposed null-space-based SWIPT system. A detailed analysis shows that the costs of dedicated EBs outweigh their benefits, ultimately diminishing overall system performance. Thus, dedicated EBs are deemed unnecessary. Instead, IBs can effectively serve both WET and WIT, achieving superior performance.
	
    \item We consider the impact of practical EH and waveform compatibility, observing that while dedicated EBs designed for specific waveforms may enhance WET, the use of IBs often provides comparable
or superior results in terms of system efficiency.
	
	\item Moreover, we design a low computational complexity algorithm to address the proposed problem, leveraging the fact that the power consumption for WET is significantly higher than for WIT. This observation allows us to disregard the benefits of WIT for WET, simplifying the entire null-space-based system to a point-to-point model.
	
	\item Numerical results validate our analysis. Although the use of dedicated EBs reduces received radio frequency (RF) power by $0.2-1$ dB, the direct current (DC) power surpasses that of the counterpart by 1 dB when $M=64$ and $P_{\max}=15$, demonstrating the effectiveness of dedicated EBs. Moreover, the low computational complexity algorithm reduces the complexity by 91.43\% and 98.54\% when $M=8,K^I=K^E=2$ and $M=16, K^E=K^I=4$, respectively, with negligible performance loss.
\end{itemize}

The organization of this paper is as follows: Section \ref{sec:II} provides an overview of the system model, detailing the signal model, practical energy harvesting model, and various waveform structures. Section \ref{sec:III} discusses the principles of null-space-based SWIPT, including beamforming techniques and designs with low computational complexity. Numerical results are presented in Section \ref{sec:IV}, followed by conclusions in Section \ref{sec:V}.

\emph{Notation:} $\mathbf{I}_{M\times M}$ denotes the $M \times M$ identity matrix, $[\cdot]_i$ refers to the $i$-th element of a vector, while $[\cdot]_{i,j}$ denotes the $(i,j)$-th element of a matrix. The imaginary unit is represented as $\mathbbm{i} = \sqrt{-1}$. The Euclidean norm is indicated by $||\cdot||$, and the absolute value by $|\cdot|$. The function $\text{diag}(\cdot)$ constructs a diagonal matrix. The operators $(\cdot)^{T}$ and $(\cdot)^{H}$ denote the transpose and conjugate transpose, respectively. Finally, $\mathcal{CN}$ denotes the circularly symmetric complex Gaussian distribution.

\section{System Model}\label{sec:II}
In this section, we describe the system model and architecture for the proposed null-space-based SWIPT system.  As depicted in Fig. \ref{fig:systemmodel}, a hybrid access point (HAP) equipped with $M$ antennas as a uniform linear array (ULA) and serves for $K^I$ single antenna IUs and $K^E$ single antenna energy users (EUs) simultaneously, and denoted as $\boldsymbol{\mathcal{K}}^I=\left[\mathcal{K}_I^1,\cdots, \mathcal{K}_I^{K^I}\right]$ and $\boldsymbol{\mathcal{K}}^E=\left[\mathcal{K}_E^1,\cdots, \mathcal{K}_E^{K^E}\right]$, respectively. We assume $M\geq K^I+K^E$ without loss of generality. Note that each EU is connected to an energy harvester (EH) for RF-to-DC energy conversion. And in this model, WET benefits from both EBs and IBs, while IUs experience interference from both intra-WIT and inter-WET, i.e., the $i$-th IU experiences interferences from $l$-th IB, $\forall l\in \boldsymbol{\mathcal{K}}^I,l \neq i$ and $j$-th EB, $\forall j\in \boldsymbol{\mathcal{K}}^E$.

\subsection{Signal Model}
Quasi-static block fading Rician channels are taken into account. Specifically, the channels between HAP-IUs and HAP-EUs are denoted by $\mathbf{H}_I=\left[\mathbf{h}_I^1,\cdots,\mathbf{h}_I^{K^I}\right]^H\in\mathbb{C}^{K^I\times M}$ and $\mathbf{H}_E=\left[\mathbf{h}_E^1,\cdots,\mathbf{h}_E^{K^E}\right]^H\in\mathbb{C}^{K^E\times M}$, respectively. And we have
\begin{align}
&\mathbf{h}_I^i=\sqrt{\frac{\kappa_{I}^i}{1+\kappa_{I}^i}}{\bar{\mathbf{h}}}^i_{I}+\sqrt{\frac{1}{1+\kappa_{I}^i}}{\tilde{\mathbf{h}}}^i_{I}, \forall i \in \boldsymbol{\mathcal{K}}^I,\\
&\mathbf{h}_E^j=\sqrt{\frac{\kappa_{E}^j}{1+\kappa_{E}^j}}{\bar{\mathbf{h}}}^j_{E}+\sqrt{\frac{1}{1+\kappa_{E}^j}}{\tilde{\mathbf{h}}}^j_{E}, \forall j \in \boldsymbol{\mathcal{K}}^E,
\end{align}
where $\mathbf{h}_I^i\in\mathbb{C}^{M\times 1}$ and $\mathbf{h}_E^j\in\mathbb{C}^{M\times 1}$. $\kappa_I^i$ and $\kappa_E^j$ denote the Rician factor of HAP-$i$-th IUs and HAP-$j$-th EUs channel. $\bar{\mathbf{h}}^i_{I}\in\mathbb{C}^{M\times 1}$ and ${\bar{\mathbf{h}}}^j_{E}\in\mathbb{C}^{M\times 1}$ denote the line-of-sight (LoS) component of Rician channel, while ${\tilde{\mathbf{h}}}^i_{I}\in\mathbb{C}^{M\times 1}$ and ${\tilde{\mathbf{h}}}^j_{E}\in\mathbb{C}^{M\times 1}$ denote the non-line-of-sight (NLoS) component. And ${\tilde{\mathbf{h}}}^i_{I}$ and ${\tilde{\mathbf{h}}}^j_{E}$ are both circularly symmetric complex white Gaussian (CSCG) and
we have ${\tilde{\mathbf{h}}}^i_{I}\sim\mathcal{CN}(\mathbf{0}, \mathbf{I}_{M\times M})$ and ${\tilde{\mathbf{h}}}^j_{E}\sim\mathcal{CN}(\mathbf{0}, \mathbf{I}_{M\times M})$.

Note that the HAP arrange as a ULA, thus, the LoS component can be further expressed as $\left[\bar{\mathbf{h}}^i_{I}\right]_m=\frac{1}{\sqrt{M}}e^{(m-1)\mathbbm{i}\varpi_i}, m\in M$, where $\varpi_i=2\pi d\sin{\phi^i_\text{dep}}/\lambda=\pi\sin{\phi^i_\text{dep}}$ by setting $d/\lambda=1/2$ without loss of generality, and $\phi^i_{\text{dep}}$ denotes the angle of departure (AOD) from the HAP to the $i$-th IUs/EUs.

Denoted the information signal and energy signal for the $i$-th IU and $j$-th EU by $s_I^i, \forall i\in\boldsymbol{\mathcal{K}}^I$ and $s_E^j, \forall j\in \boldsymbol{\mathcal{K}}^E$, and the signal transmitted by the HAP can be expressed as
\begin{align}
	\mathbf{x}=\sum_{i\in\boldsymbol{\mathcal{K}^I}}\mathbf{w}_is_I^i+\sum_{j\in\boldsymbol{\mathcal{K}}^E}\mathbf{v}_js_E^j,
\end{align}
where $\mathbf{w}_i\in\mathbb{C}^{M\times 1}$ and $\mathbf{v}_j\in\mathbb{C}^{M\times 1}$ denote the IB and the EB for the $i$-th IU and $j$-th EU, respectively.

We assume that the information signal $s_I^i, \forall i\in\boldsymbol{\mathcal{K}}^I$ is an independent and identically distributed (i.i.d) CSCG random variable with zero mean and unit variance. Regarding the energy signal $s_E^j, \forall j\in\boldsymbol{\mathcal{K}}^E$, we consider both Gaussian and deterministic sinusoidal waveforms, which will be elaborated upon in detail in the following sections. Thus, the signal-to-interference-plus-noise-ratio (SINR) of the $k$-th IU can be expressed as
\begin{align}
	&\gamma_k\nonumber\\
	&=\frac{\left|\left(\mathbf{h}_I^k\right)^H\mathbf{w}_k\right|^2}{\sum\limits_{i\in \boldsymbol{\mathcal{K}}^I,i\neq k}\left|\left(\mathbf{h}_I^k\right)^H\mathbf{w}_i\right|^2+\sum\limits_{j\in \boldsymbol{\mathcal{K}}^E}\left|\left(\mathbf{h}_I^k\right)^H\mathbf{v}_js_E^j\right|^2+\frac{\sigma^2_0}{\varrho_{H2I,k}}},\label{eqn:sinr}
\end{align}
where $\varrho_{H2I,k}$ denotes the path-loss from the HAP to the $k$-th IU, and Eq. \eqref{eqn:sinr} is derived from that $s_I^j$ is an i.i.d CSCG variable, $\sigma_0$ denotes the noise power. Similarly, the received RF power at the $l$-th EU can be expressed as
\begin{align}
	E^\text{RF}_l=&\sum_{i\in \boldsymbol{\mathcal{K}}^I}\varrho_{H2E,l}\left|\left(\mathbf{h}_E^l\right)^H\mathbf{w}_i\right|^2\nonumber\\
	&\qquad\qquad\qquad+\sum_{j\in \boldsymbol{\mathcal{K}^E}}\varrho_{H2E,l}\left|\left(\mathbf{h}_E^l\right)^H\mathbf{v}_js_E^j\right|^2,
\end{align}
where $\varrho_{H2E,j}$ denotes the path-loss from the HAP to the $j$-th EU.

\subsection{Practical Energy Harvester}\label{sec:practical_EH}

The practical non-linear EH model proposed in \cite{NonLinearEH} captures the dynamics of RF energy conversion efficiency at different input power levels. Specifically, the total harvested energy at $j$-th EU can be expressed as
\begin{align}
	f(P) &= \frac{M_s}{X(1+\exp(-a(P-b)))}-Y, \quad\left[\text{Watt}\right]\nonumber\\
	X &= \frac{\exp(ab)}{1+\exp(ab)}, \quad Y= \frac{M_s}{\exp(ab)},
\end{align}
where $f(\cdot)$ denotes the non-linear relationship between the input RF power and harvested energy, $P$ is the input RF power. $a$ and $b$ denote the joint impact of the resistances, the capacitances, and the circuit sensitivity on the rectifying
process. Note that we assume that each EU equipped with a single antenna, which is connected to a non-linear EH.

\subsection{Different Waveforms}
Many studies have examined the impact of waveforms design on SWIPT systems\cite{Buro_sine, waveform_YQD}. To capture two representative extremes of communication waveforms, we focus on Gaussian waveforms and deterministic sinusoidal waveforms. The Gaussian waveforms corresponds to communication signals under very high-order QAM modulation schemes, such as 4096-QAM, and serves as a model for highly randomized signal structures. In contrast, the deterministic sinusoidal waveforms represent low-order constant-envelope modulation schemes, such as BPSK and QPSK, thereby providing a benchmark for structured and energy-efficient signal designs.
\begin{itemize}
	\item Gaussian Waveform: let the energy signal $s_E^j\sim\mathcal{CN}\left(0,1\right),\forall j\in\boldsymbol{\mathcal{K}}^E$. Thus, the received RF power by the $l$-th EU can be expressed as
	\begin{align}
		&E^{\text{GW}}_l=\sum_{i\in \boldsymbol{\mathcal{K}}^I}\varrho_{H2E,l}\left|\left(\mathbf{h}_E^l\right)^H\mathbf{w}_i\right|^2\nonumber\\
		&\qquad\qquad\qquad+\sum_{j\in \boldsymbol{\mathcal{K}^E}}\varrho_{H2E,l}\left|\left(\mathbf{h}_E^l\right)^H\mathbf{v}_js_E^j\right|^2.\label{eqn:GW}
	\end{align}
	Note that since $s_E^j\sim\mathcal{CN}\left(0,1\right),\forall j\in\boldsymbol{\mathcal{K}}^E$, we have $\mathbb{E}\left(\left|s_E^j(s_E^j)^H\right|\right)=1$.
	\item Deterministic Sinusoidal Waveform\cite{waveform_YQD}: similar with the Gaussian waveform, the received power with deterministic sinusoidal waveform can be expressed as
	\begin{align}
		&E^{\text{DSW}}_l=\sum_{i\in \boldsymbol{\mathcal{K}}^I}\varrho_{H2E,l}\left|\left(\mathbf{h}_E^l\right)^H\mathbf{w}_i\right|^2\nonumber\\
		&\qquad\qquad\qquad\qquad+\sum_{j\in \boldsymbol{\mathcal{K}^E}}\varrho_{H2E,l}\left|\left(\mathbf{h}_E^l\right)^H\mathbf{v}_j\right|^2,\label{eqn:DSW}
	\end{align}
	since $||s_E^j||^2=1$.
\end{itemize}

\begin{Rem}\label{remark:1}
	Note that although both the Gaussian waveform or deterministic sinusoidal waveform have the similar expression as Eq. \eqref{eqn:GW} and Eq. \eqref{eqn:DSW}, we have $||s_E^j||^2=1$ when the deterministic sinusoidal waveform is adopted. As for the Gaussian waveform, we have $\mathbb{E}\left(\left|s_E^j(s_E^j)^H\right|\right)=1$. Moreover, the energy waveform can be considered as any information signal from the $K^I$ IUs, meaning that the EUs and IUs may share the same waveform and beamforming simultaneously. But for the deterministic sinusoidal waveform, we have to adopt dedicated EBs for energy transfer, since the difference between $s_I^i,\forall i\in\boldsymbol{\mathcal{K}}^I$ and $s_E^j,\forall j\in\boldsymbol{\mathcal{K}}^E$.
\end{Rem}

\subsection{Problem formulation}
In this paper, we aim to maximize the total harvested energy of all EUs while satisfying the communication demands of the IUs. The problem can be formulated as
\begin{align}
\text{(P1): } &\max_{\mathbf{w}_i,\mathbf{v}_j} \sum_{l\in\boldsymbol{\mathcal{K}}^E}f\left(E^{\text{GW/DSW}}_l\right)\label{eqn:P1}\\
\text{s.t. }\
&\log_2\left(1+\gamma_k\right)>C_\text{thre},\forall k\in \boldsymbol{\mathcal{K}}^I,\label{eqn:P1Ca} \tag{\ref{eqn:P1}a}\\
&\sum_{i\in \boldsymbol{\mathcal{K}}^I}\left|\left|\mathbf{w}_i\right|\right|^2+\sum_{j\in \boldsymbol{\mathcal{K}}^E}\left|\left|\mathbf{v}_j\right|\right|^2\leq P_\text{max},\label{eqn:P1Cb} \tag{\ref{eqn:P1}b}
\end{align}
where $C_\text{thre}$ represents the channel capacity constraints for all IUs, and $P_{\text{max}}$ denotes the maximum transmit power. Note that, due to the variations in energy waveforms, we may obtain diverse results, such as different power allocations for EBs and IBs, different waveform selections, and different transmission strategies.

\section{Null-Space-based SWIPT}\label{sec:III}
In this section, we first analyze the transmission characteristics of the SWIPT system and formulate a common optimization problem for both types of waveforms. Then, we propose a null-space-based SWIPT algorithm for interference suppression. Based on the proposed algorithm, we derive several important recommendations. Finally, we simplify the algorithm for a deterministic sinusoidal waveform according to the power levels of WET and WIT.

With the aid of first-order Taylor expansion, the objective function can be reformulated as\cite{waveform_YQD}
\begin{align}
	\sum_{l\in\boldsymbol{\mathcal{K}}^E} f\left(E_l^\text{GW/DSW}\right)
	\approx  \sum_{l\in\boldsymbol{\mathcal{K}}^E} E_l^\text{GW/DSW}+K^Ep_0,\label{eqn:firstorderTaylor}
\end{align}
where Eq. \eqref{eqn:firstorderTaylor} comes from the first-order Taylor expansion and $p_0$ is a constant in Taylor expansion.

For the Gaussian waveform, we have
\begin{align}
	\mathbb{E}\left(f\left(E^{\text{GW}}_l\right)\right)=&\sum_{i\in \boldsymbol{\mathcal{K}}^I}\varrho_{H2E,l}\left|\left(\mathbf{h}_E^l\right)^H\mathbf{w}_i\right|^2\nonumber\\
	&+\sum_{j\in \boldsymbol{\mathcal{K}^E}}\varrho_{H2E,l}\left|\left(\mathbf{h}_E^l\right)^H\mathbf{v}_j\right|^2+K^Ep_0\label{eqn:GW_mean},
\end{align}

Observe from Eq. \eqref{eqn:DSW} and Eq. \eqref{eqn:GW_mean}, we can rewritten objective $\text{(P1)}$ equivalently as
\begin{align}
	\text{(P2): } &\max_{\mathbf{w}_i,\mathbf{v}_j}\sum_{l\in \boldsymbol{\mathcal{K}}^E}\left(\sum_{i\in \boldsymbol{\mathcal{K}}^I}\varrho_{H2E,l}\left|\left(\mathbf{h}_E^l\right)^H\mathbf{w}_i\right|^2\right.\nonumber\\
	&\qquad\qquad\qquad\quad\left.+\sum_{j\in \boldsymbol{\mathcal{K}}^E}\varrho_{H2E,l}\left|\left(\mathbf{h}_E^l\right)^H\mathbf{v}_j\right|^2\right)\label{eqn:P2}\\
	\text{s.t.} \
	&\log_2\left(1+\gamma_k\right)>C_\text{thre},\forall k\in \boldsymbol{\mathcal{K}}^I,\label{eqn:P2Ca} \tag{\ref{eqn:P2}a}\\
	&\sum_{i\in \boldsymbol{\mathcal{K}}^I}\left|\left|\mathbf{w}_i\right|\right|^2+\sum_{j\in \boldsymbol{\mathcal{K}}^E}\left|\left|\mathbf{v}_j\right|\right|^2\leq P_\text{max},\label{eqn:P2Cb} \tag{\ref{eqn:P2}b}
\end{align}
where objective $\text{(P2)}$ is to maximize the sum-power transferred to all EUs. Constraint \eqref{eqn:P2Ca} is the SINR constraint and constraint \eqref{eqn:P2Cb} is the total transmit power limitation.

\begin{Rem} \label{remark:2}
Note that Eq. \eqref{eqn:P2} can be regarded as a unified expression that maximizes harvested energy with both type of waveforms. It concurrently addresses the symbol-averaged\footnote{Symbol-averaged means that, within a single coherence interval and for a fixed channel realization, the objective is optimized as an expectation over the Gaussian distribution (i.e., $\mathbb{E}\left(\left|s_E^j(s_E^j)^H\right|\right)=1$). By contrast, the deterministic sinusoid admits instantaneous (per-symbol) optimization with $||s_E^j||^2=1$.} optimization with Gaussian waveform, i.e., shown in Eq. \eqref{eqn:GW_mean} and the instantaneous optimization with deterministic sinusoidal waveform, i.e., shown in Eq. \eqref{eqn:DSW}. Objective $\text{(P2)}$ enables us to temporarily disregard the influence of energy signals, thereby allowing us to concentrate on the design of beamforming and the allocation of power for both IUs and EUs. Moreover, as emphasized in Remark \ref{remark:1}, the difference between deterministic sinusoidal waveform and information signal (i.e., Gaussian waveform) prevent the transmission of energy waveform loading in the IBs, which means that we need dedicated EBs for energy transfer while deterministic sinusoidal waveform is adopted.
\end{Rem}

\subsection{Null Space-based SWIPT with Gaussian Waveform}

\begin{figure}
	\centering
	\includegraphics[width=0.7\linewidth]{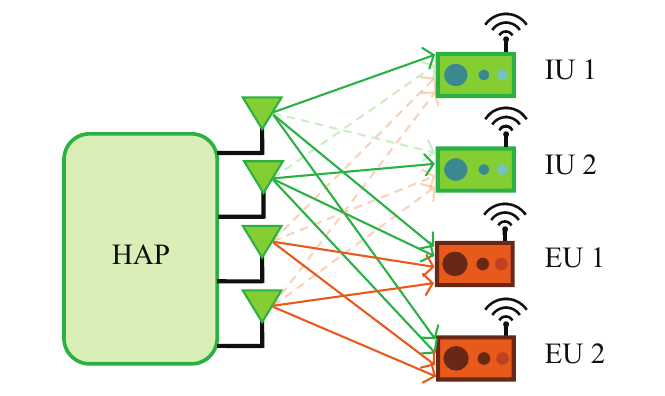}
	\caption{An illustration of the proposed system, where the EUs benefit from the IBs.}
	\label{fig:nullspace1}
\end{figure}

In this section, we propose a null-space-based algorithm to solve the problem $\text{(P2)}$. Specifically, the null-space-based beamforming is used here for inter-IUs and EUs interference elimination, which can be expressed as
\begin{align}
	\mathbf{H}_I^{\boldsymbol{\mathcal{K}}^I\backslash \mathcal{K}_I^i}
	\mathbf{w}_i=\mathbf{0}_{\left(K^I-1\right)\times1},\forall i\in \boldsymbol{\mathcal{K}}^I,\label{eqn:elimi_IBs}\\
	\mathbf{H}_I
	\mathbf{v}_j=\mathbf{0}_{K^I\times1},\forall j\in \boldsymbol{\mathcal{K}}^E\label{eqn:elimi_EBs},
\end{align}
where $\mathbf{H}_I^{\boldsymbol{\mathcal{K}}^I\backslash \mathcal{K}_I^i}=\left[\mathbf{h}_I^1,\cdots,\mathbf{h}_I^{i-1}, \mathbf{h}_I^{i+1}, \cdots\mathbf{h}_I^{K^I}\right]^H\in\mathbb{C}^{\left(K^I-1\right)\times M}$. Note that Eq. \eqref{eqn:elimi_IBs} and Eq. \eqref{eqn:elimi_EBs} eliminate the interference from other IBs and EBs for IUs, respectively. Thus, we construct the null space as
\begin{align}
	&\mathbf{N}^I_{i}=\text{null}\left(\mathbf{H}_I^{\boldsymbol{\mathcal{K}}^I\backslash \mathcal{K}_I^i}\right)\in\mathbb{C}^{ M\times\left(M-r^I\right)},\label{eqn:nullNI}\\
	&\mathbf{N}^E_{j}=\text{null}\left(\mathbf{H}_I\right)\in\mathbb{C}^{ M\times\left(M-r^E\right)},\label{eqn:nullNE}
\end{align}
where $M-r^I$ and $M-r^E$ denote the available column of the null space. In practically, we can obtain the null space matrix via singular value decomposition (SVD). Specifically, we can decompose $\mathbf{H}_I\in\mathbb{C}^{M\times K^I}$ by SVD as
\begin{align}
	\mathbf{H}_I&=\mathbf{U} \boldsymbol{\Sigma} \mathbf{V}^{\mathrm{H}}\nonumber\\
	&=\mathbf{U}\left[\begin{array}{ll}
	\boldsymbol{\Sigma}_1 & \mathbf{0}_{K^I\times \left(M-K^I\right)}
	\end{array}\right]\left[\begin{array}{ll}
	\mathbf{V}_1 & \mathbf{V}_0
	\end{array}\right]^{\mathrm{H}},\label{eqn:SVDdecompose}
\end{align}
where $\mathbf{U}\in\mathbb{C}^{K^I\times K^I}$, $\boldsymbol{\Sigma}_1\in\mathbb{C}^{K^I\times K^I}$, $\mathbf{V}_1\in\mathbb{C}^{M\times r^E}$ and $\mathbf{V}_0\in\mathbb{C}^{M\times(M-r^E)}$. The variable $r^E$ denotes the available column of the null space, and we have $1\leq r^E\leq M-1$. We have
\begin{align}
	\text{null}\left(\mathbf{H}_I\right)=\mathbf{V}_0,
\end{align}

\begin{Lemma}\label{lemma:1}
	The optimal $r^I=\text{row}\left\{\mathbf{H}_I^{\boldsymbol{\mathcal{K}}^I\backslash \mathcal{K}_I^i}\right\}=K^I-1$ and $r^E=\text{row}\left\{\mathbf{H}_I\right\}=K^I$.
\end{Lemma}
\begin{IEEEproof}
	Please refer to Appendix \ref{app:A} for detailed proof.
\end{IEEEproof}

Obtain the null space, we can construct the EBs and IBs as
\begin{align}
	\bar{\mathbf{w}}_i=\mathbf{N}^I_{i}\mathbf{b}_i, \forall i\in \boldsymbol{\mathcal{K}}^I,\label{eqn:bar_w}\\
	\bar{\mathbf{v}}_j=\mathbf{N}_j^E\mathbf{d}_j, \forall j\in \boldsymbol{\mathcal{K}}^E,\label{eqn:bar_v}
\end{align}
where $\mathbf{b}_i\in\mathbb{C}^{(M-r^I)\times 1}$ and $\mathbf{d}_j\in\mathbb{C}^{(M-r^E)\times 1}$ denote the auxiliary vector for $i$-th IU and $j$-th EU, respectively. $\bar{\mathbf{w}}_i\in\mathbb{C}^{M\times 1}$ and $\bar{\mathbf{v}}_j\in\mathbb{C}^{M\times 1}$ denote the normalized IBs and EBs, which can be expressed as
\begin{align}
	\mathbf{w}_i=\sqrt{P_I^i}\bar{\mathbf{w}}_i,\label{eqn:w}\\
	\mathbf{v}_j=\sqrt{P_E^j}\bar{\mathbf{v}}_j,\label{eqn:v}
\end{align}
where $P_I^i$ and $P_E^j$ denote the transmit power of $i$-th IB and $j$-th EB, respectively.

With the aid of null space, we can further simplified and study the problem $\text{(P2)}$ by applying the technique of semidefinite relaxation (SDR) as
\begin{align}
	\text{(P2.1): } &\max_{\mathbf{W}_i, \mathbf{V}}\sum_{i\in \boldsymbol{\mathcal{K}}^I}\text{trace}\left(\mathbf{S}\mathbf{W}_i\right)+\text{trace}\left(\mathbf{S}\mathbf{V}\right)\label{eqn:P21}\\
	\text{s.t.} \
	&\text{trace}\left(\mathbf{h}_I^i\left(\mathbf{h}_I^i\right)^H\mathbf{W}_k\right)>\frac{\left(2^{C_\text{thre}}-1\right)\sigma_0^2}{\varrho_{H2I,k}},\forall k\in \boldsymbol{\mathcal{K}}^I,\label{eqn:P21Ca} \tag{\ref{eqn:P21}a}\\
	&\sum_{i\in \boldsymbol{\mathcal{K}}^I}\text{trace}\left(\mathbf{W}_i\right)+\text{trace}\left(\mathbf{V}\right)\leq P_\text{max},\label{eqn:P21Cb} \tag{\ref{eqn:P21}b}\\
	&\mathbf{W}_i\succeq\mathbf{0},\forall i\in \boldsymbol{\mathcal{K}}^I,\quad \mathbf{V}\succeq\mathbf{0}.\label{eqn:P21Cc} \tag{\ref{eqn:P21}c}
\end{align}
where $\mathbf{S}=\sum_{j\in \boldsymbol{\mathcal{K}^E}}\mathbf{h}_E^j\left(\mathbf{h}_E^j\right)^H$, $\mathbf{W}_i=\mathbf{w}_i\mathbf{w}_i^H$ and $\mathbf{V} = \sum_{j\in\boldsymbol{\mathcal{K}}^E}\mathbf{v}_j\mathbf{v}_j^H$. Objective \eqref{eqn:P21} is to maximize the sum-power transferred to all EUs, and \eqref{eqn:P21Ca} and \eqref{eqn:P21Cb} represent the channel capacity constraint and total transmit power constraint, respectively. Note that with the aid of null space, the interference signals are completely eliminated, which can be expressed as
\begin{align}
	&\text{trace}\left(\mathbf{h}_{I}^{k}\left(\mathbf{h}_{I}^{k}\right)^H\mathbf{W}_i\right)=0,\forall i\in\boldsymbol{\mathcal{K}}^I,i\neq k, \forall k\in \boldsymbol{\mathcal{K}}^I,\nonumber\\
	&\text{trace}\left(\mathbf{h}_{I}^{k}\left(\mathbf{h}_{I}^{k}\right)^H\mathbf{V}\right)=0,\forall k\in \boldsymbol{\mathcal{K}}^I,
\end{align}
resulting in the expression in \eqref{eqn:P21Ca}. Fig. \ref{fig:nullspace1} illustrates the null-space-based SWIPT scheme, in which intra-WIT and inter-WET interference in WIT is effectively suppressed, and WET benefits not only from EBs but also from IBs.

Observe from Eq. \eqref{eqn:bar_w}-\eqref{eqn:v}, we can rewrite the received power by $k$-th EU as
\begin{align}
	&\sum_{i\in \boldsymbol{\mathcal{K}}^I}\left|\left(\mathbf{h}_E^k\right)^H\mathbf{w}_i\right|^2+\sum_{j\in \boldsymbol{\mathcal{K}^E}}\left|\left(\mathbf{h}_E^k\right)^H\mathbf{v}_j\right|^2\nonumber\\
	=&\sum_{i\in \boldsymbol{\mathcal{K}}^I}\left|\left(\mathbf{h}_E^k\right)^H\mathbf{N}^I_{i}\sqrt{P_I^i}\mathbf{b}_i\right|^2+\sum_{j\in \boldsymbol{\mathcal{K}^E}}\left|\left(\mathbf{h}_E^k\right)^H\mathbf{N}^E_{j}\sqrt{P_E^j}\mathbf{d}_j\right|^2\nonumber\\
	=&\sum_{i\in \boldsymbol{\mathcal{K}}^I}\left|\mathbf{h}_{E,I}^{k,i}\sqrt{P_I^i}\mathbf{b}_i\right|^2+\sum_{j\in \boldsymbol{\mathcal{K}^E}}\left|\mathbf{h}_{E,E}^{k,j}\sqrt{P_E^j}\mathbf{d}_j\right|^2\nonumber\\
	=&\sum_{i\in \boldsymbol{\mathcal{K}}^I}\left|\mathbf{h}_{E,I}^{k,i}\mathbf{b}^\S_i\right|^2+\sum_{j\in \boldsymbol{\mathcal{K}^E}}\left|\mathbf{h}_{E,E}^{k,j}\mathbf{d}^\S_j\right|^2,\label{eqn:eqniv}
\end{align}
where the equivalent channel $\left(\mathbf{h}_{E,I}^{k,i}\right)^H=\left(\mathbf{h}_E^k\right)^H\mathbf{N}^I_{i}$, $\left(\mathbf{h}_{E,E}^{k,j}\right)^H=\left(\mathbf{h}_E^k\right)^H\mathbf{N}^E_{j}$, $\mathbf{b}_i^\S=\sqrt{P_I^i}\mathbf{b}_i$ and $\mathbf{d}_j^\S=\sqrt{P_E^j}\mathbf{d}_j$.

Thus, we can further rewrite problem $\text{(P2.1)}$ as
\begin{align}
	\text{(P2.2):  } &\max_{\mathbf{w}_i,\mathbf{v}_j}\sum_{i\in \boldsymbol{\mathcal{K}}^I}\text{trace}\left(\mathbf{S}_{E,i}\mathbf{B}_i\right)+\text{trace}\left(\mathbf{S}_{E}\mathbf{D}\right)\label{eqn:P22}\\
	\text{s.t. }\
	&\text{trace}\left(\mathbf{h}_{I,I}^{k}\left(\mathbf{h}_{I,I}^{k}\right)^H\mathbf{B}_k\right)>\frac{\left(2^{C_\text{thre}}-1\right)\sigma_0^2}{\varrho_{H2I,k}},\forall k\in \boldsymbol{\mathcal{K}}^I,\label{eqn:P22Ca} \tag{\ref{eqn:P22}a}\\
	&\sum_{i\in \boldsymbol{\mathcal{K}}^I}\text{trace}\left(\mathbf{B}_i\right)+\text{trace}\left(\mathbf{D}\right)\leq P_\text{max},\label{eqn:P22Cb} \tag{\ref{eqn:P22}b}\\
	&\mathbf{B}_i\succeq\mathbf{0},\forall i\in \boldsymbol{\mathcal{K}}^I,\quad \mathbf{D}\succeq\mathbf{0}.\label{eqn:P22Cc} \tag{\ref{eqn:P22}c}
\end{align}
where $\mathbf{S}_{E,i}=\sum_{j\in\boldsymbol{\mathcal{K}}^E}\mathbf{h}_{E,I}^{j,i}\left(\mathbf{h}_{E,I}^{j,i}\right)^H$, $\mathbf{S}_E=\sum_{k\in\boldsymbol{\mathcal{K}^E}}\mathbf{h}_{E,E}^{k,j}\left(\mathbf{h}_{E,E}^{k,j}\right)^H=\sum_{k\in\boldsymbol{\mathcal{K}^E}}\mathbf{h}_{E,E}^{k}\left(\mathbf{h}_{E,E}^{k}\right)^H$ since all the EUs share the same null space, i.e., $\mathbf{N}^E_j=\mathbf{N}^E,\forall j\in \boldsymbol{\mathcal{K}}^E$. And $\left(\mathbf{h}_{I, I}^k\right)^H=\left(\mathbf{h}_I^k\right)^H\mathbf{N}^I_{k}$. $\mathbf{B}_i=\mathbf{b}^\S_i\left(\mathbf{b}^\S_i\right)^H\in\mathbb{C}^{\left(M-r^I\right)\times \left(M-r^I\right)}$, $\mathbf{D}=\sum_{j\in\boldsymbol{\mathcal{K}}^E}\mathbf{d}^\S_j\left(\mathbf{d}^\S_j\right)^H\in\mathbb{C}^{\left(M-r^E\right)\times \left(M-r^E\right)}$.

\begin{Lemma}\label{lemma:2}
	The optimal solution of $\mathbf{B}_i$ and $\mathbf{D}$ satisfy: $\text{rank}\left(\mathbf{B}_i\right)=1,\forall i\in\boldsymbol{\mathcal{K}}^I$ and $\mathbf{D}=\mathbf{0}$.
\end{Lemma}
\begin{IEEEproof}
	Please refer to Appendix \ref{app:B} for detailed proof.
\end{IEEEproof}

According to Lemma \ref{lemma:2}, the problem $\text{(P2.2)}$ can be further simplified as
	\begin{align}
		\text{(P2.3):  } &\max_{\mathbf{B}_i}\sum_{i\in \boldsymbol{\mathcal{K}}^I}\text{trace}\left(\mathbf{S}_{E,i}\mathbf{B}_i\right)\label{eqn:P23}\\
		\text{s.t. } \
		&\text{trace}\left(\mathbf{h}_{I,I}^{k}\left(\mathbf{h}_{I,I}^{k}\right)^H\mathbf{B}_k\right)>\frac{\left(2^{C_\text{thre}}-1\right)\sigma_0^2}{\varrho_{H2I,k}},\forall k\in \boldsymbol{\mathcal{K}}^I,\label{eqn:P23Ca} \tag{\ref{eqn:P23}a}\\
		&\sum_{i\in \boldsymbol{\mathcal{K}}^I}\text{trace}\left(\mathbf{B}_i\right)\leq P_\text{max},\label{eqn:P23Cb} \tag{\ref{eqn:P23}b}\\
		&\mathbf{B}_i\succeq\mathbf{0},\forall i\in \boldsymbol{\mathcal{K}}^I,\label{eqn:P23Cc} \tag{\ref{eqn:P23}c}
	\end{align}
which is a convex optimization problem, and can be solved by standard optimization method, such as the interior point method. The details of this algorithm are summarized in Algorithm \ref{alg:1}.

\begin{Rem}\label{remark:3}
	Lemma \ref{lemma:2} demonstrates that it is unnecessary to design a dedicated EB for WET using a Gaussian waveform in our proposed null-space-based system. Instead, we need only to load the energy for WET into IBs\footnote{It is important to note that all conclusions in this work are established under the assumption that the underlying optimization problem is feasible. For instance, the transmit power must be sufficiently large to at least satisfy the WIT requirements. In general, a feasibility check of the original problem is necessary to ensure solvability\cite{JieXu_NoEB}. However, such details are omitted in this paper due to space limitations.}. Note that many studies such as \cite{JieXu_NoEB} claimed that WET benefits from the dedicated EB as the interference is canceled, which is completely contrary to our conclusion, as these studies do not consider the cost of interference elimination, effectively assuming it to be free. In fact, as illustrates in Lemma \ref{lemma:2}, the gain obtained from dedicated EB is less than the cost of interference elimination, indicating that there is no need for a dedicated EB. Therefore, we arrive at a general conclusion: unless the EBs can provide additional benefits (e.g., supporting dedicated WET waveforms as discussed in Section \ref{sec:iiib}), the design of dedicated EBs inevitably leads to performance degradation. This conclusion holds broadly in the context of SWIPT problems modeled as maximizing the total harvested energy, such as problem $\text{(P2)}$, and remains valid across different scenarios and under various channel conditions.
\end{Rem}

\begin{algorithm}[t]
	\small
	\caption{Proposed null-space-based SWIPT algorithm with Gaussian waveform.}
	\begin{algorithmic}[1]\label{alg:1}
		\REQUIRE~\ HAP-to-IUs channels $\mathbf{H}_I$, HAP-to-EUs channels $\mathbf{H}_E$, $M$, $C_\text{thre}$.
		\ENSURE~\ Optimal solution $\mathbf{w}_i,\forall i\in \boldsymbol{\mathcal{K}}^I$.
		\STATE Obtain the information null space $\mathbf{N}^I_i,\forall i\in \boldsymbol{\mathcal{K}}^I$ from Eq. \eqref{eqn:nullNI}.
		\STATE Obtain the equivalent channel $\mathbf{h}_{E,I}^{k,i}$, $\mathbf{h}_{E,E}^{k,j}$, $\mathbf{h}_{I,I}^{k}$, and the matrix $\mathbf{S}_{E,i}$ and $\mathbf{S}_{E}$ from Eq. \eqref{eqn:eqniv}-\eqref{eqn:P22}.
		\STATE Solve the problem $\text{(P2.3)}$ to obtain $\mathbf{B}_i,\forall i\in \boldsymbol{\mathcal{K}}^I=\mathbf{b}^\S_i\left(\mathbf{b}^\S_i\right)^H$.
		\STATE Obtain $\mathbf{b}^\S_i$ by eigendecomposition of $\mathbf{B}_i$.
		\STATE Reconstruct $\mathbf{w}_i=\mathbf{N}^I_i\mathbf{b}^\S_i,\forall i\in \boldsymbol{\mathcal{K}}^I$.
	\end{algorithmic}
\end{algorithm}

\subsection{Null Space-based SWIPT with Deterministic Sinusoidal Waveform}\label{sec:iiib}
	In this section, we introduce the null-space-based SWIPT system with deterministic sinusoidal waveform. Note that as we mentioned in Lemma \ref{lemma:1}, the difference between information waveform $s_I^i,\forall i\in\boldsymbol{\mathcal{K}}^I$ and the deterministic sinusoidal waveform prevent adopting IBs for WET, which means that we need to design dedicated EBs for WET.

	It should be emphasized that, for analytical tractability, we approximate the nonlinear EH function in Section \ref{sec:III} using a first-order Taylor expansion, which inevitably discards the higher-order terms. However, the nonlinear behavior of EH circuits is fundamentally characterized by such higher-order moments\cite{Bruno_intro}. As a result, the first-order approximation cannot distinguish the performance differences of various waveforms passing through the nonlinear EH circuit, and solving the original problem \text{(P2.2)} therefore always yields the conclusion $\mathbf{D}=\mathbf{0}$. To compensate for this limitation, we introduce an additional factor $\eta$ to restore the waveform-induced gain inherent in the nonlinear EH model, thereby enabling a solution with $\mathbf{D}\neq\mathbf{0}$. This treatment represents a deliberate trade-off between modeling accuracy and analytical complexity.

\begin{Lemma}\label{lemma:3}
	To obtain the dedicated EB for EUs, the problem $\text{(P2.2)}$ can be rewritten as
	\begin{align}
		\text{(P2.4): } &\max_{\mathbf{w}_i,\mathbf{v}_j}\sum_{i\in \boldsymbol{\mathcal{K}}^I}\text{trace}\left(\mathbf{S}_{E,i}\mathbf{B}_i\right)+\eta\text{trace}\left(\mathbf{S}_{E}\mathbf{D}\right)\label{eqn:P24}\\
		\text{s.t. }\
		&\text{trace}\left(\mathbf{h}_{I,I}^{k}\left(\mathbf{h}_{I,I}^{k}\right)^H\mathbf{B}_k\right)>\frac{\left(2^{C_\text{thre}}-1\right)\sigma_0^2}{\varrho_{H2I,k}},\forall k\in \boldsymbol{\mathcal{K}}^I,\label{eqn:P24Ca} \tag{\ref{eqn:P24}a}\\
		&\sum_{i\in \boldsymbol{\mathcal{K}}^I}\text{trace}\left(\mathbf{B}_i\right)+\text{trace}\left(\mathbf{D}\right)\leq P_\text{max},\label{eqn:P24Cb} \tag{\ref{eqn:P24}b}\\
		&\mathbf{B}_i\succeq\mathbf{0},\forall i\in \boldsymbol{\mathcal{K}}^I,\quad \mathbf{D}\succeq\mathbf{0},\label{eqn:P24Cc} \tag{\ref{eqn:P24}c}
	\end{align}
	where $\eta$ denotes the reward factor, which motivates to allocate power for EB. And $\eta>\frac{\max_{i\in\boldsymbol{\mathcal{K}}^I}(\xi^{\max}_{E,i})}{\xi^{\max}_E}+\delta$, where $\xi^{\max}_{E,i}$ and $\xi^{\max}_E$ denote the dominant eigenvalue of matrix $\mathbf{S}_{E,i}$ and $\mathbf{S}_E$, respectively. And the definition of $\delta$ can be found in Appendix \ref{app:C}.
\end{Lemma}
\begin{IEEEproof}
	Please refer to Appendix \ref{app:C} for detailed proof.
\end{IEEEproof}

\begin{Rem}\label{remark:4}
From Lemma \ref{lemma:2}, it is evident that dedicated EB reduces overall energy performance, as the cost of mitigating energy interference surpasses the benefits gained from dedicated EB. Thus, we obtain $\mathbf{D} = \mathbf{0}$. In contrast, Lemma \ref{lemma:3} introduces the reward factor $\eta$, which assumes sufficient benefits from the energy waveform, leading to power allocation for dedicated EB and $\mathbf{D} \neq \mathbf{0}$ when $\eta > \frac{\max_{i\in\boldsymbol{\mathcal{K}}^I}(\xi^{\max}_{E,i})}{\xi^{\max}_E}+\delta$. Moreover, since calculating $\delta$ can be cumbersome, we can simplify the process by setting $\delta$ to a sufficiently large value, e.g., $\delta = 10$, to avoid complex computations.
\end{Rem}

Problem $\text{(P2.4)}$ is a convex optimization problem, and can be solved by standard optimization method, such as the interior point method.

Note that the discrepancies between the information waveform $s_I^i, \forall i \in \boldsymbol{\mathcal{K}}^I$, and the deterministic sinusoidal waveform preclude the joint optimization of IBs and EB for maximal harvested energy. Thus, the solution of $\text{(P2.4)}$, strictly speaking, is suboptimal. Additionally, EUs typically require much higher received power than IUs \cite{JieXu_NoEB, subopt1, subopt2, subopt3}, which motivates us to obtain the IBs and EBs by disregarding the benefits of IBs to EUs, thereby achieving a low-complexity design for the proposed system.
\begin{figure}
	\centering
	\includegraphics[width=0.7\linewidth]{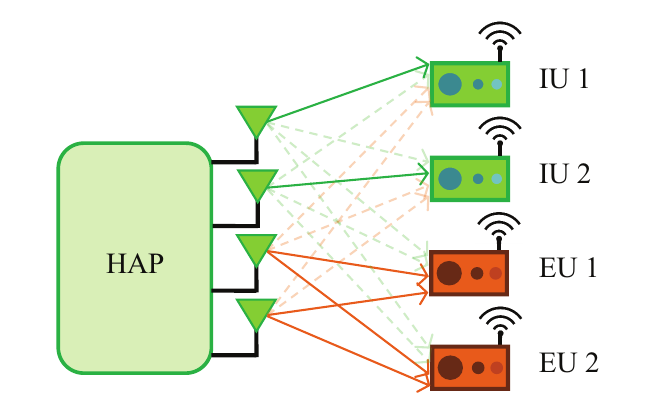}
	\caption{An illustration of the proposed system. In this scenario, we focus solely on the EBs for the EUs, disregarding any potential benefits from the IBs.}
	\label{fig:nullspace2}
\end{figure}

\begin{table*}[t]
	\centering
	\caption{The Complexity of Different Algorithm.}
	\begin{tabular}{c|c}
	\toprule
	\textbf{Method} & \textbf{Computational Complexity}\\
	\hline
	\textbf{Algorithm. \ref{alg:1}} &  $\mathcal{O}\left(
		\left(K^I\right)^{3.5}\left(M-r^I\right)^{3.5}+
		K^I\left(M-r^I\right)^3+
		K^IK^EM^3+
		K^IM\left(K^I-1\right)^2-K^IK^EM^2r^I\right)$\\
	\hline
	\textbf{Problem $\text{(P2.4)}$}& \parbox{10cm}{$\mathcal{O}\left(
		\left(K^I\right)^{3.5}\left(M-r^I\right)^{3.5}+
		\left(M-r^E\right)^{3.5}+
		K^I\left(M-r^I\right)^3+\left(M-r^E\right)^3
		+K^IK^EM^3\right.$
		$\left.\qquad\qquad\qquad\qquad\qquad\qquad\qquad\qquad
		+K^IM\left(K^I-1\right)^2+M\left(K^I\right)^2
		-K^IK^EM^2r^I
		\right)$} \\
	\hline
	\textbf{Algorithm. \ref{alg:2}}& $\mathcal{O}\left(
		\left(M-r^E\right)^3+
		K^IM^2\left(M-r^I\right)+
		K^IM\left(K^I-1\right)^2+M\left(K^I\right)^2
		\right)$ \\
	\hline
	\textbf{Benchmark}& $\mathcal{O}\left(
		\left(\left(K^I\right)^{3.5}+1\right)M^{3.5}+\left(K^I+1\right)M^3
		\right)$\\
	\hline
	\textbf{Benchmark w/o $\mathbf{V}$}& $\mathcal{O}\left(
		\left(K^I\right)^{3.5}M^{3.5}+K^IM^3
		\right)$ \\
	\bottomrule
	\end{tabular} \label{table:complexityofal}
\end{table*}

As shown in Fig. \ref{fig:nullspace2}, by disregarding the contribution of IBs at the EUs and leveraging the null-space interference suppression scheme, the SWIPT design is decoupled into two independent subproblems: a point-to-point WIT problem and a WET problem. Specifically, to obtain a low-complexity design of the proposed system, we first divided the original problem $\text{(P2)}$ into two sub-problems, which can be expressed as
\begin{align}
	\text{(Sub.1):}\ &\min_{\mathbf{w}_i} \ \sum_{i\in\boldsymbol{\mathcal{K}}^I}\left|\left|\mathbf{w}_i\right|\right|^2\label{eqn:sub1}\\
	\text{s.t.}\
	&\log_2\left(1+\frac{\varrho_{H2I,k}\left|\left(\mathbf{h}_I^k\right)^H\mathbf{w}_k\right|^2}{\sigma^2_0}\right)>C_\text{thre},\forall k\in \boldsymbol{\mathcal{K}}^I. \tag{\ref{eqn:sub1}a}\label{eqn:subSINR}
\end{align}
Denote by $\left|\left|\mathbf{w}_i\right|\right|^2=P_I^i,\forall i\in\boldsymbol{\mathcal{K}}^I$ the transmit power of $i$-th IU, we have

\begin{align}
	\text{(Sub.2):}\ &\max_{\mathbf{V}=\mathbf{v}\mathbf{v}^H} \ \text{trace}\left(\mathbf{S}\mathbf{V}\right)\label{eqn:sub2}\\
	\text{s.t. }\
	&\text{trace}\left(\mathbf{V}\right)\leq P_{\max}-\sum_{i\in\boldsymbol{\mathcal{K}}^I}P_I^i.\tag{\ref{eqn:sub2}a}
\end{align}

Problem $\text{(Sub.1)}$ aims to minimize the transmit power for WIT, while the remaining power is allocated for WET to maximize the harvested power in problem $\text{(Sub.2)}$. By utilizing the null space, we can eliminate interference signals, thereby deriving constraint \eqref{eqn:subSINR}.

It is worth emphasizing that although the solutions to $\text{(Sub.1)}$ and $\text{(Sub.2)}$ are suboptimal, the performance loss compared to the optimal solution is limited and can even be considered negligible. This is because the power levels of IBs and EBs often differ by tens of dB \cite{JieXu_NoEB, subopt1, subopt2, subopt3}.

Observe from the $\text{(Sub.1)}$ that, the problem is simplified as a point-to-point communication problem as the interferences are eliminated shown in constraint \eqref{eqn:subSINR}. Thus, we aim to maximize the $i$-th target signal as
\begin{align}
	\text{(Sub.1.1)}& \ \max_{\mathbf{b}_i} \ P_I^i\left|\left(\mathbf{h}_I^i\right)^H\bar{\mathbf{w}}_i\right|^2=P_I^i\left|\left(\mathbf{h}_I^i\right)^H\mathbf{N}^I_i\mathbf{b}_i\right|^2\label{eqn:sub11}\\
	\text{s.t. }\
	&\quad \left|\left|\mathbf{b}_i\right|\right|^2=1,\tag{\ref{eqn:sub11}a}\label{eqn:sub11a}
\end{align}
where the constraint \eqref{eqn:sub11a} is come from
\begin{align}
	\left|\left|\bar{\mathbf{w}}_i\right|\right|^2&=\left|\left|\mathbf{N}^I_i\mathbf{b}_i\right|\right|^2=\text{trace}\left(\mathbf{N}^I_i\mathbf{b}_i\mathbf{b}_i^H\left(\mathbf{N}^I_i\right)^H\right)\nonumber\\
	&=\text{trace}\left(\left(\mathbf{N}^I_i\right)^H\mathbf{N}^I_i\mathbf{b}_i\mathbf{b}_i^H\right)\nonumber\\
	&=\text{trace}\left(\mathbf{I}_{M-r^I_i}\mathbf{B}_i\right)\nonumber\\
	&=\left|\left|\mathbf{b}_i\right|\right|^2,
\end{align}

Thus, the maximum-ratio transmission (MRT) can be adopted for the simplified point-to-point problem, as
\begin{align}
	&\mathbf{b}_i=\frac{\left(\mathbf{N}^I_i\right)^H\mathbf{h}_I^i}{\left|\left|\left(\mathbf{N}^I_i\right)^H\mathbf{h}_I^i\right|\right|},
    &\bar{\mathbf{w}}_i=\frac{\mathbf{N}^I_i\left(\mathbf{N}^I_i\right)^H\mathbf{h}_I^i}{\left|\left|\left(\mathbf{N}^I_i\right)^H\mathbf{h}_I^i\right|\right|},\forall i\in\boldsymbol{\mathcal{K}}^I,\label{eqn:bar_w_sub}
\end{align}

and
\begin{align}
	&\beta_i=\frac{\text{trace}\left(\mathbf{h}_I^i\left(\mathbf{h}_I^i\right)^H\mathbf{N}_i^I\left(\mathbf{N}_i^I\right)^H\mathbf{h}_I^i\left(\mathbf{h}_I^i\right)^H\mathbf{N}_i^I\left(\mathbf{N}_i^I\right)^H\right)}{\left|\left|\left(\mathbf{N}_i^I\right)^H\mathbf{h}_I^i\right|\right|^2},\nonumber\\
	&P_I^i=\frac{\left(2^{C_\text{thre}}-1\right)\sigma_0^2}{\varrho_{H2I}^i\beta_i}.\label{eqn:power_sub}
\end{align}

Benefiting from the closed-form MRT solution rather than solving the optimization problem $\text{(Sub.1)}$, the complexity of obtaining $\bar{\mathbf{w}}_i$ is substantially reduced. Accordingly, the final beamforming vector can be expressed as $\mathbf{w}_i=\sqrt{P_I^i}\bar{\mathbf{w}_i},\forall i\in\boldsymbol{\mathcal{K}}^I$, where $P_I^i$ denotes the allocated transmit power for the $i$-th IB.

As for $\text{(Sub.2)}$, we can rewrite as
\begin{align}
	\text{(Sub.2.1):}\ &\max_{\mathbf{D}=\mathbf{d}^{\S}\left(\mathbf{d}^{\S}\right)^H} \ \text{trace}\left(\mathbf{S}_E\mathbf{D}\right)\label{eqn:sub21}\\
	\text{s.t. }\
	&\text{trace}\left(\mathbf{D}\right)\leq P_{\max}-\sum_{i\in\boldsymbol{\mathcal{K}}^I}P_I^i.\tag{\ref{eqn:sub21}a}
\end{align}

The normalized vector $\mathbf{d}=\mathbf{d}^\S/ ||\mathbf{d}^\S||$ is aligned with the eigenvector of $\mathbf{S}_E$ corresponding to the maximal eigenvalue, as mentioned above. $P_E=P_{\max}-\sum_{i\in\boldsymbol{\mathcal{K}}^I}P_I^i$ denotes the remaining power for problem \text{(Sub.2.1)} and finally we can obtain $\mathbf{v} = \sqrt{P_E}\mathbf{N}^E\mathbf{d}$. The details of this algorithm are summarized in Algorithm \ref{alg:2}.

\begin{algorithm}[t]
	\small
	\caption{Proposed null-space-based SWIPT algorithm with deterministic sinusoidal waveform.}
	\begin{algorithmic}[1]\label{alg:2}
		\REQUIRE~\ HAP-to-IUs channels $\mathbf{H}_I$, HAP-to-EUs channels $\mathbf{H}_E$, $M$, $C_\text{thre}$.
		\ENSURE~\ Optimal solution  $\mathbf{w}_i,\forall i\in \boldsymbol{\mathcal{K}}^I$, $\mathbf{v}$.
		\STATE Obtain the information null space $\mathbf{N}^I_i,\forall i\in \boldsymbol{\mathcal{K}}^I$ and energy null space $\mathbf{N}^E$ from Eq. \eqref{eqn:nullNI}-\eqref{eqn:nullNE}.
		\STATE Obtain $\mathbf{w}_i=\sqrt{P_I^i}\bar{\mathbf{w}_i},\forall i\in\boldsymbol{\mathcal{K}}^I$, where $P_I^i$ and $\bar{\mathbf{w}_i}$ is come from Eq. \eqref{eqn:bar_w_sub}-\eqref{eqn:power_sub}.
		\STATE Obtain the normalized vector $\mathbf{d}$ from eigendecomposition of $\mathbf{S}_E$.
		\STATE Reconstruct $\mathbf{v}=\sqrt{P_E}\mathbf{N}^E\mathbf{d}$, where $P_E=P_{\max}-\sum_{i\in\boldsymbol{\mathcal{K}}^I}P_I^i$.
	\end{algorithmic}
\end{algorithm}

\subsection{Computational Complexity}

Regarding Algorithm \ref{alg:1}, the complexity of computing the information null space requires $K^I$ instances of SVD, each of which can be performed with a complexity of $\mathcal{O}\left(M(K^I-1)^2\right)$ \cite[Lecture 31]{trefethen2022numerical}, assuming that $M \geq K^I + K^E > K^I - 1$. The complexity of calculating the equivalent channels $\mathbf{h}_{E,I}^{k,i}$, $\mathbf{h}_{E,E}^{k,j}$, $\mathbf{h}_{I,I}^{k}$, and the matrices $\mathbf{S}_{E,i}$ and $\mathbf{S}_{E}$ are primarily depend on the complexity of $\mathbf{S}_{E,i}$, which can be approximated as $\mathcal{O}\left(K^IK^E \left(M^3 - M^2r^I\right)\right)$. Solving problem $\text{(P2.3)}$ to obtain $\mathbf{b}^\S$ has a complexity of $\mathcal{O}\left(\left(K^I(M - r^I)\right)^{3.5} + K^I(M - r^I)^3\right)$ when using the interior-point method, resulting in the total complexity as shown in Alg. \ref{alg:1} of Table \ref{table:complexityofal}.

The complexity of solving $\text{(P2.4)}$, including null space and equivalent channel calculations, is $\mathcal{O}\left(K^I M (K^I - 1)^2 + M (K^I)^2 + K^IK^E \left(M^3 - M^2r^I\right)\right)$. Solving the problem using the interior-point method has a complexity of $\mathcal{O}\left(\left(K^I (M - r^I)\right)^{3.5} + \left(M - r^E\right)^{3.5}\right)$. Additionally, the complexity of performing eigendecomposition to obtain $\mathbf{b}^\S_i$ and $\mathbf{d}$ is $\mathcal{O}\left(K^I (M - r^I)^3 + (M - r^E)^3\right)$ \cite[Lecture 25]{trefethen2022numerical}. The total complexity is summarized in $\text{(P2.4)}$ of Table. \ref{table:complexityofal}.

Regarding Algorithm \ref{alg:2}, the complexity of null space calculation is $\mathcal{O}\left(K^I M (K^I - 1)^2 + M (K^I)^2\right)$. The complexity to obtain the IBs via MRT is $\mathcal{O}\left(K^I M^2 (M - r^I)\right)$. Moreover, the final eigendecomposition to obtain $\mathbf{v}$ requires $\mathcal{O}\left((M - r^E)^3\right)$. The total complexity is summarized in Alg. \ref{alg:2} of Table \ref{table:complexityofal}.

\section{Numerical Results}\label{sec:IV}
This section provides a performance evaluation of the proposed systems under specified configurations. The HAP-to-EU and HAP-to-IU distances are set to 5 meters (m) and 50 m, respectively, with reference signal attenuation at 1 m of 30 dB. Path loss exponents for HAP-EU and HAP-IU channels are 2.2 and 3.2, respectively, and noise power, $\sigma_0^2$, is -84 dBm. Non-linear EH parameters are configured as $a=150$, $b=0.024$, and $M_s=24$ mW \cite{EH_value}. Unless otherwise noted, default parameters are $M=16$, $K^I = K^E = 2$, and $P_{\max} = 2$.
\begin{figure}
	\centering
	\includegraphics[width=1\linewidth]{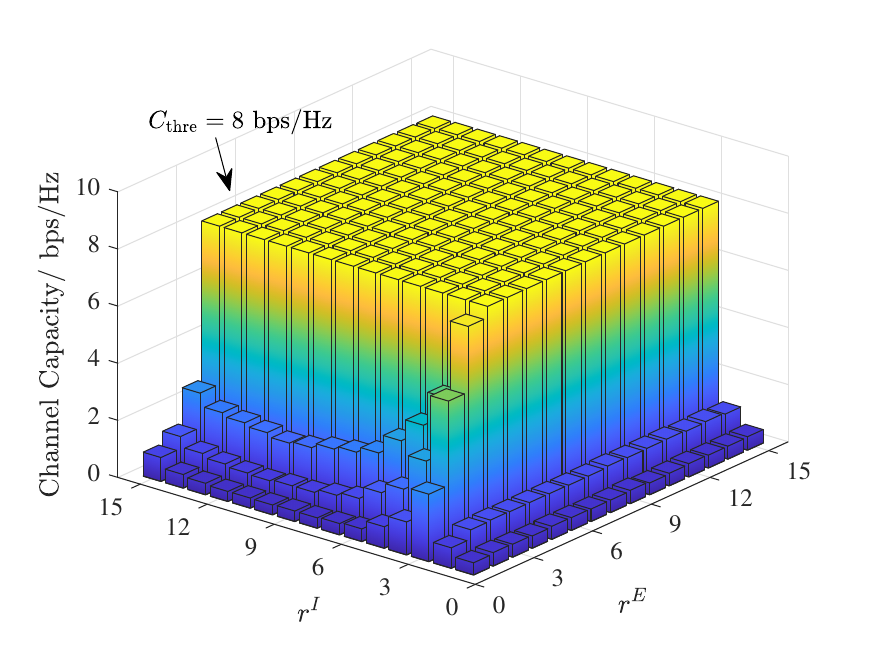}
    \caption{Worst-case channel capacity among $K^I$ IUs for different $r^I$ and $r^E$ values, with $P_{\max}=2$ W transmit power.}
	\label{fig:exp1-1}
\end{figure}
\begin{figure}
	\centering
	\includegraphics[width=1\linewidth]{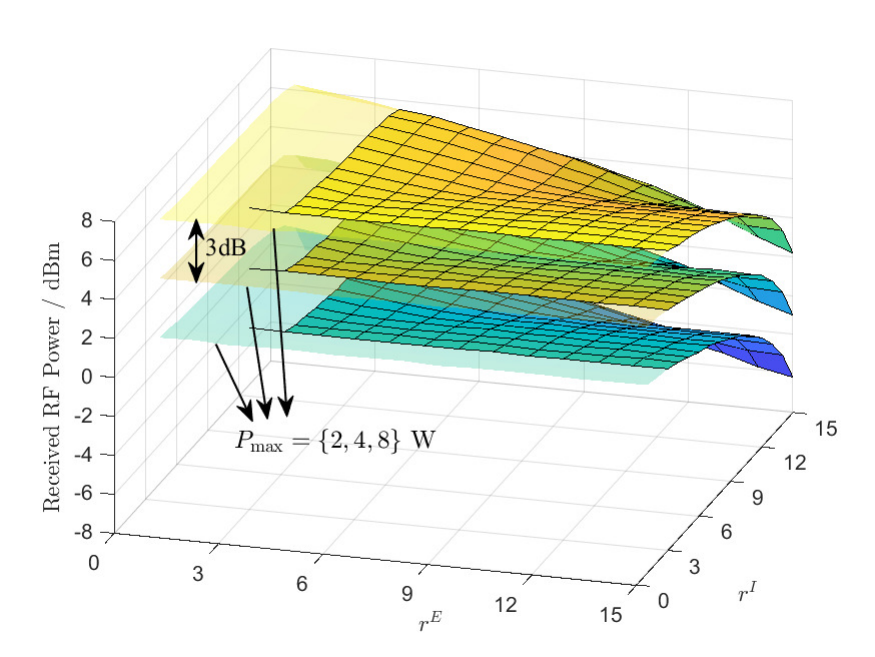}
    \caption{Received RF power across different $r^I$ and $r^E$ values, with a 3 dB gain for $P_{\max} = 2$, 4, and 8 W.}
	\label{fig:exp1}
\end{figure}
For simplicity, we initially omit non-linear EH, focusing instead on the received RF power variation with parameter adjustments.

Fig. \ref{fig:exp1-1} and Fig. \ref{fig:exp1}  depict the worst-case channel capacity for $K^I$ IUs and the total RF power received by $K^E$ EUs, derived by solving problem $\text{(P2.2)}$ for different $r^I$ and $r^E$ with $K^E=K^I=4$ and $M=16$. For cases with $r^E\geq K^I=4$ and $r^I\geq K^I-1=3$, the channel capacity reaches 8 bps/Hz, satisfying the quality of service (QoS) constraints in \eqref{eqn:P22Ca}. The translucent portions in Fig. \ref{fig:exp1} correspond to scenarios where QoS requirements are unmet, rendering these data points invalid. For valid cases, as both $r^I$ and $r^E$ increase, received power decreases. When $r^E > r^I$, forming the lower triangular region from bottom-left to top-right in Fig. \ref{fig:exp1}, the received power remains constant with variations in $r^E$. In contrast, when $r^E \leq r^I$, the received power varies with changes in $r^E$. This behavior occurs because, for the cases $r^E > r^I$, dedicated EB is not required according to Lemma \ref{lemma:2}, making $r^E$ irrelevant. However, for the cases $r^E < r^I$, the Cauchy interlacing theorem (see Appendix \ref{app:B}) does not hold, necessitating dedicated EB for EUs, and $r^E$ thus influences the received power, as also discussed in Remark \ref{remark:1}. Consequently, unless otherwise specified in the following section, we set $r^I = K^I - 1$ and $r^E = K^I$.
\begin{figure}
	\centering
	\includegraphics[width=0.9\linewidth]{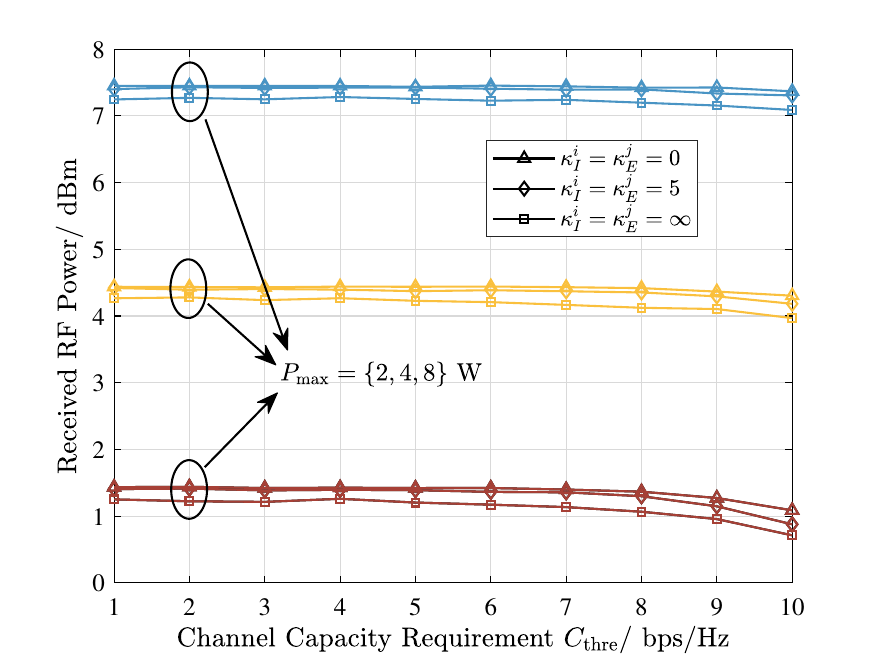}
	\caption{Received RF energy under varying channel capacity requirements.}
	\label{fig:exp2}
\end{figure}

Fig. \ref{fig:exp2} illustrates received power under various channel capacity demands, determined by solving problem $\text{(P2.3)}$. As capacity requirements increase, received power decreases, highlighting a WIT-WET trade-off. A high Rician factor ($\kappa_I^i=\kappa_E^j=\infty, \forall i\in \boldsymbol{\mathcal{K}}^I, j\in \boldsymbol{\mathcal{K}}^E$) degrades performance due to low-rank channels, reducing spatial multiplexing gains and increasing interference and power needs for WIT, consistent with findings in \cite{WQQ}. This decrease is offset by increasing transmit power, shown by comparisons with $P_{\max} = 2$ and $P_{\max} = 8$.

\begin{figure}
	\centering
	\includegraphics[width=0.9\linewidth]{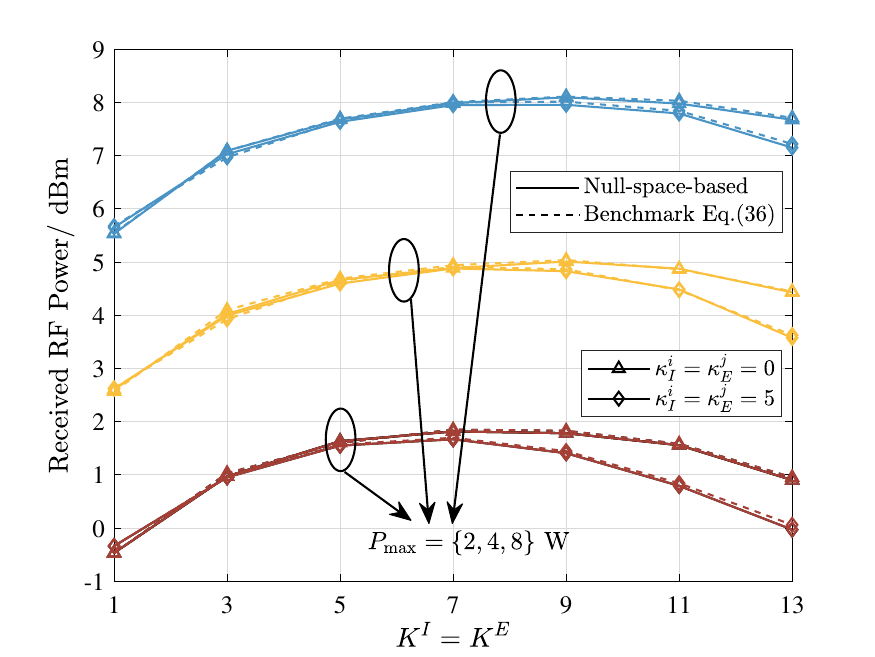}
	\caption{Received RF power under different IU and EU counts.}
	\label{fig:exp3}
\end{figure}

To examine the received power performance of the proposed null-space-based system, we introduce a benchmark that directly solves the SDR problem as follows:
\begin{align}
	&\text{(Benchmark): }\max_{\mathbf{W}_i,\mathbf{V}} \sum_{i\in \boldsymbol{\mathcal{K}}^I}\mathrm{trace}(\mathbf{S}\mathbf{W}_i)+\mathrm{trace}(\mathbf{S}\mathbf{V})\label{eqn:trace_simp}\\
	&\text{s.t. } \frac{\mathrm{trace}\left(\mathbf{h}^i_I\left(\mathbf{h}^i_I\right)^H\mathbf{W}_i\right)}{\left(2^{C_\text{thre}}-1\right)}-\mathrm{trace}\left(\mathbf{h}^i_I\left(\mathbf{h}^i_I\right)^H\mathbf{V}\right)-\frac{\sigma_0^2}{\varrho_{H2I,i}}\nonumber\\
	&\qquad-\sum_{l\in K^I,l\neq i}\mathrm{trace}\left(\mathbf{h}^i_I\left(\mathbf{h}^i_I\right)^H\mathbf{W}_l\right)\geq 0,\forall i\in \boldsymbol{\mathcal{K}}^I,\tag{\ref{eqn:trace_simp}a}\label{eqn:trace_simpCa}\\
	&\,\quad \sum_{i\in \boldsymbol{\mathcal{K}}^I}\mathrm{trace}\left(\mathbf{W}_i\right)+\mathrm{trace}\left(\mathbf{V}\right)\leq P_\text{max},\tag{\ref{eqn:trace_simp}b}\label{eqn:trace_simpCb}\\
	&\,\,\quad \mathbf{W}_i\succeq \mathbf{0},\forall i\in \boldsymbol{\mathcal{K}}^I, \mathbf{V} \succeq\mathbf{0}.\tag{\ref{eqn:trace_simp}c}\label{eqn:trace_simpCc}
\end{align}
This convex optimization problem can be efficiently solved using standard methods, such as the interior point method.

Fig. \ref{fig:exp3} shows received RF power for varying IU and EU counts, indicating an initial increase followed by a decrease across $P_{\max} = {2, 4, 8}$ W. Initially, adding IUs and EUs increases total RF power as more EUs utilize WET. However, beyond a certain count, power decreases due to reduced degrees of freedom. Scenarios with $\kappa_I^i=\kappa_E^j=5$ show fewer degrees of freedom than Rayleigh fading channels ($\kappa_I^i=\kappa_E^j=0$), leading to an earlier decline in RF power. This demonstrates that the null-space-based scheme will significantly benefits from MIMO systems and large-scale antenna technologies. In addition, compared with the benchmark scheme in Eq. \eqref{eqn:trace_simp}, the proposed null-space-based approach exhibits the same variation trend, which further confirms that the performance degradation is attributed to the change in spatial degrees of freedom rather than to the null-space algorithm itself.

\begin{figure}
	\centering
	\includegraphics[width=0.9\linewidth]{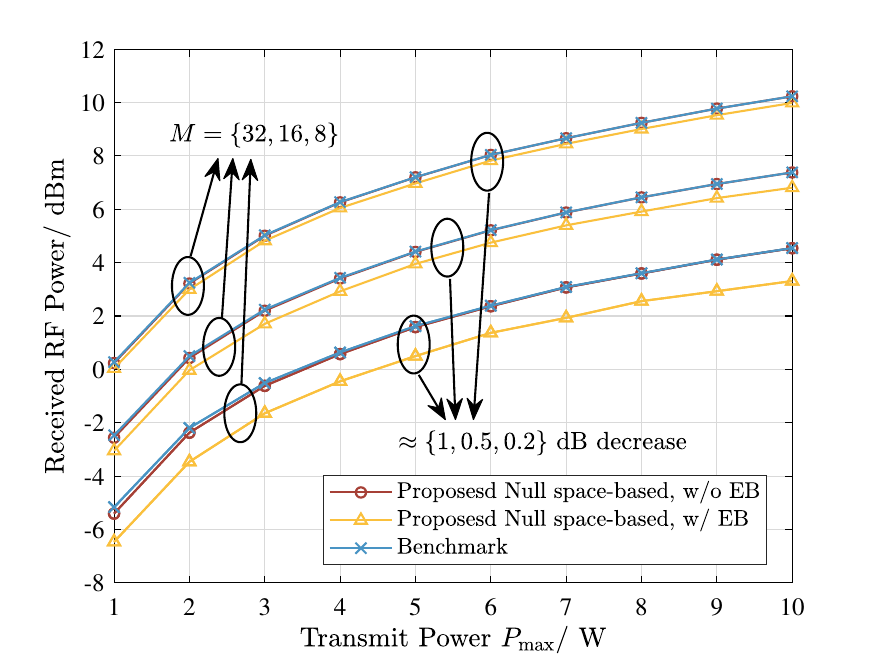}
	\caption{Received RF energy with different communication requirements. The dedicated EB will decrease the energy performance as mentioned in Lemma \ref{lemma:2} and Remark \ref{remark:3}.}
	\label{fig:exp4}
\end{figure}

Fig. \ref{fig:exp4} shows received RF energy variation with transmit power. It is noteworthy that the proposed null-space-based scheme, i.e., without dedicated EB, performs comparably to the benchmark. In contrast, the null-space-based scheme with dedicated EB shows decreases of 1 dB, 0.5 dB, and 0.2 dB when $M = \{8, 16, 32\}$, respectively. It is important to highlight that the dedicated EB counterpart is obtained by solving problem $\text{(P2.4)}$, where the reward factor $\eta$ leads to the solution incorporating dedicated EB, as discussed in Remark \ref{remark:4}. Moreover, the limited spatial degrees of freedom (e.g., $M=8$) and transmit power (e.g., $P_{\max} = 1$) may also degrade the performance of the proposed null-space-based scheme, as illustrated in Fig. \ref{fig:exp4}.

To further assess the applicability of the proposed null-space-based algorithm, we consider the case of imperfect CSI, modeled as
\begin{align}
	\mathbf{H}_I^{e}=\sqrt{1-\rho^2}\,\mathbf{H}_I+\rho \mathbf{H}_{I,n},\nonumber\\	  
	\mathbf{H}_E^{e}=\sqrt{1-\rho^2}\,\mathbf{H}_E+\rho \mathbf{H}_{E,n},
\end{align}
where $\rho$ denotes the error level and the error vectors follow a CSCG distribution, i.e., $\mathbf{h}_n^i \sim \mathcal{CN}(0,\sigma_H^2)$. Under this model, the resulting beamforming vector $\mathbf{w}_i^e$ leads to residual interference, i.e.,$\mathbf{H}_I^{\boldsymbol{\mathcal{K}}^I\backslash \mathcal{K}_I^i}\mathbf{w}_i^e=\rho\mathbf{H}^{\boldsymbol{\mathcal{K}}^I\backslash \mathcal{K}_I^i}_{I,n} \mathbf{w}_i^e \neq \mathbf{0}$, which cannot be fully eliminated. Since only the statistical properties of channel errors are available, advanced robust methods (e.g., the General S-Procedure\cite{9180053,boyd1994linear}) are required for rigorous design. In this paper, we restrict attention to numerical results to illustrate possible performance degradation, while robust extensions are left for future research.

\begin{figure}[!t]
	\centering
	{
		\subfigure[Capacity.]{\includegraphics[width = 0.45\linewidth]{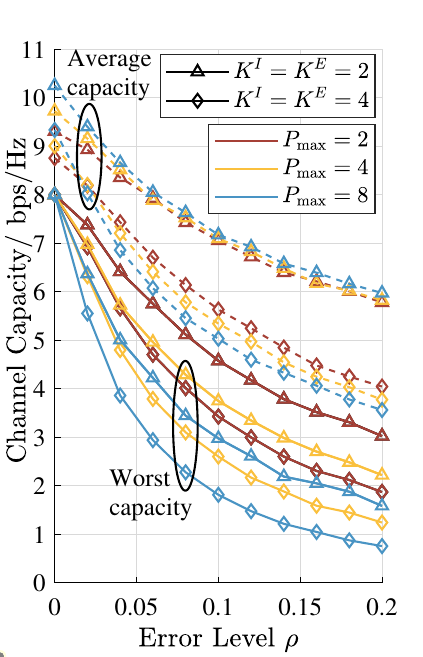}}
		\hspace{-0.4cm} 
		\subfigure[RF power.]{\includegraphics[width = 0.45\linewidth]{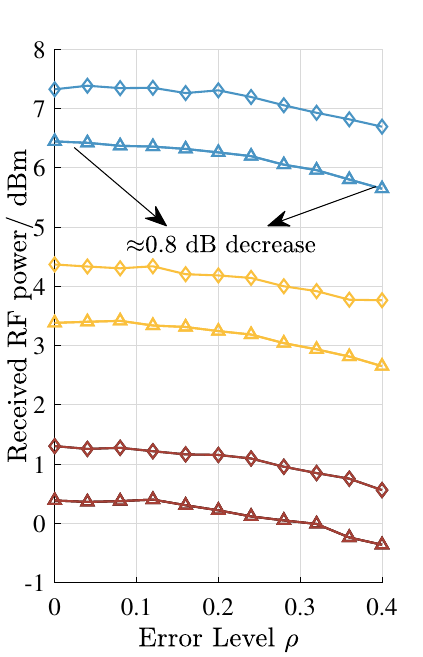}}
	}
	\setlength{\abovecaptionskip}{0pt}
	\setlength{\belowcaptionskip}{0pt}
	\caption{The SWIPT performance with imperfect CSI.}
	\label{fig:errorSWIPT}
\end{figure}

As illustrated in Fig. \ref{fig:errorSWIPT}, increasing CSI errors lead to a reduction in both the worst-case channel capacity and the total harvested RF power, with the degradation in capacity becoming more pronounced when the number of IUs/EUs increases. Nevertheless, the average capacity across multiple IUs decreases more gradually, since part of the WET power is inherently superimposed on certain IBs in problem $\text{(P2.3)}$, thereby improving the average performance. Moreover, it is observed that the robustness of channel capacity to CSI imperfections diminishes as the transmit power $P_\text{max}$ grows. This is because the interference leakage resulting from CSI mismatch scales proportionally with $P_\text{max}$, while the useful signal component is effectively constrained by the SINR requirement or the beamforming gain. Consequently, the SINR deteriorates more rapidly, leading to a sharper decline in channel capacity performance. In contrast, received RF power exhibits comparatively stronger robustness against CSI errors.

\begin{figure}
	\centering
	\includegraphics[width=0.9\linewidth]{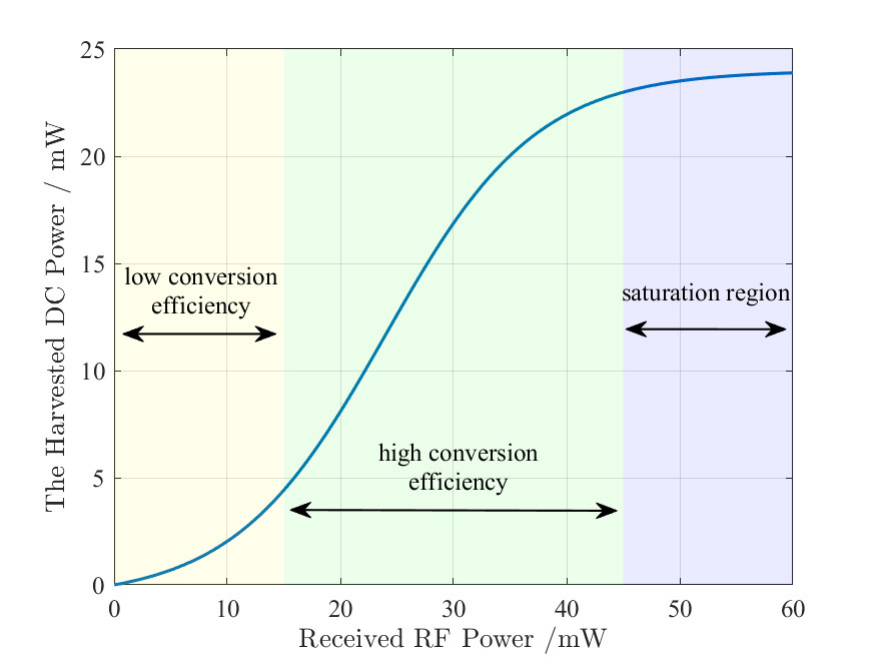}
	\caption{The RF-DC conversion efficiency of EH with different input RF power.}
	\label{fig:exp_EH_eff}
\end{figure}

Take the non-linear EH introduced in Section \ref{sec:practical_EH} into consideration. We first illustrates the characteristics of non-linear EH in Fig. \ref{fig:exp_EH_eff}. Note that the RF-DC conversion efficiency first increasing with the received RF power increasing, and then decreasing when approach saturation region. This non-linear characteristic may effect the harvested power when we adopt different waveforms.

\begin{figure}
	\centering
	\includegraphics[width=0.9\linewidth]{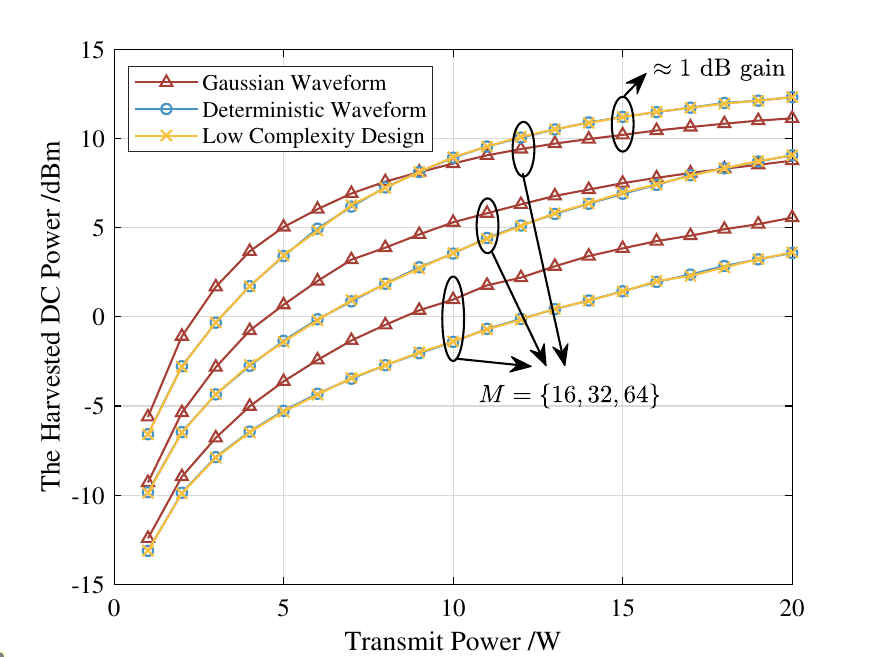}
	\caption{Harvested DC power with different transmite power. }
	\label{fig:exp5_waveformEH}
\end{figure}

In Fig. \ref{fig:exp5_waveformEH}, the harvested DC power shows that while both Gaussian and deterministic sinusoidal waveforms see an increase with rising transmit power, the Gaussian waveform outperforms the deterministic one over the range $P_{\max}=1\to 18$ for $M=32$ and $P_{\max}=1\to9$ for $M=64$. This disparity becomes more evident as antenna count and transmit power rise. The reasons for this trend can be summarized as follows:
\begin{enumerate}
	\item \textbf{Efficiency Gain with High Amplitude}:  as $M$ and $P_{\max}$ increase, the input RF power transitions from the low conversion efficiency region to the high conversion efficiency region of the non-linear EH as shown in Fig. \ref{fig:exp_EH_eff}, significantly increasing the harvested power. Moreover, the Gaussian waveform, characterized by a higher instantaneous amplitude, reaches the high-efficiency region even at lower power levels, making it advantageous at initial transmit power increments. However, as the transmit power increases, while the input RF power of a deterministic sinusoidal waveform remains within the high conversion efficiency region, the Gaussian waveform may occasionally enter the low conversion efficiency or saturation regions, potentially reducing the harvested power performance.

	\item \textbf{Degrees of Freedom in Null Space Scheme}: with larger $M$, the degrees of freedom in the null-space-based scheme increase, which reduces the cost of dedicated EB, as shown in Fig. \ref{fig:exp4}. Consequently, as $M$ grows, the harvested power of the deterministic sinusoidal waveform begins to outperform that of the Gaussian waveform at an earlier stage.
\end{enumerate}

Notably, the difference between the proposed low-complexity design and its counterpart by solving $\text{(P2.4)}$ is negligible, which shows the effectiveness of the proposed low-complexity design. In fact, we evaluate the allocated power of both WIT and WET. Table \ref{table:MKP_results} shows that even under the scenario with $P=1$, $M=16$, and $K^I=K^E=4$, the power allocated to WET is still 9.2 dB higher than that for WIT. In most other cases, the WET power allocation exceeds that of WIT by more than 13 dB. Furthermore, since the IBs must simultaneously support the WIT task, the effective ratio between the received RF power contributed by EBs and that contributed by IBs should be even larger than the WET-to-WIT allocation ratio, thereby confirming that the IBs contribution to WET can be safely neglected.

\begin{table}[t]
  \centering
  \caption{WET-to-WIT Power Allocation Ratio.}
  \label{table:MKP_results}
  \setlength{\tabcolsep}{8pt}
  \begin{tabular}{ccccc}
    \toprule
    & \multicolumn{2}{c}{$K^I=K^E=2$} & \multicolumn{2}{c}{$K^I=K^E=4$} \\
    \cmidrule(lr){2-3}\cmidrule(lr){4-5}
    $P_\text{max}$/ W & $M=16$& $M=32$ & $M=16$ & $M=32$ \\
    \midrule
     1 & 13.9 & 17.2 & \textbf{9.2}  & 13.4 \\
     2 & 17.0 & 20.3 & 12.5 & 16.6 \\
     3 & 18.8 & 22.1 & 14.3 & 18.5 \\
     4 & 20.0 & 23.4 & 15.7 & 19.7 \\
     5 & 21.0 & 24.3 & 16.7 & 20.6 \\
     6 & 21.8 & 25.1 & 17.5 & 21.4 \\
     7 & 22.4 & 25.7 & 18.0 & 22.1 \\
     8 & 23.0 & 26.4 & 18.8 & 22.7 \\
     9 & 23.6 & 26.8 & 19.3 & 23.1 \\
    10 & 24.0 & 27.3 & 19.8 & 23.7 \\
    \bottomrule
  \end{tabular}
  \vspace{1mm}
 \begin{minipage}{0.96\linewidth}
    \footnotesize
    $\textbf{Note}$: All entries are ratios in dB, computed as
    $10\log_{10}\left({P_E}/{\sum_{i=1}^{\boldsymbol{\mathcal{K}}^I} P_I^{i}}\right)$.
  \end{minipage}
\end{table}

To enable a fairer comparison with the benchmark, we have included in Table \ref{table:complexityofal} the computational complexity of directly solving the SDR problem, denoted as Benchmark and formulated in Eq. \eqref{eqn:trace_simp}. As shown in \cite[Proposition 3.1]{JieXu_NoEB}, the Benchmark admits $\mathbf{V}=\mathbf{0}$, which further reduces the computational complexity. This variant is referred to as Benchmark w/o $\mathbf{V}$. The complexities of both Benchmark and Benchmark w/o $\mathbf{V}$ are summarized in Table \ref{table:complexityofal}. In particular, when $M=8$, $K^E=K^I=2$, and $M=16$, $K^E=K^I=4$, the complexity of the proposed Algorithm \ref{alg:2} is reduced by 91.43\% and 98.54\% relative to the Problem (P2.4), by 94.01\% and 99.26\% relative to Benchmark, and by 93.34\% and 99.25\% relative to  Benchmark w/o $\mathbf{V}$, respectively, as reported in Table \ref{table:complexityofalReduction}. Furthermore, the performance degradation is negligible, as demonstrated in Fig. \ref{fig:exp5_waveformEH}, which further corroborates the effectiveness of the proposed low-complexity algorithm.

\begin{table}[t]
  \centering
  \caption{Complexity Reduction of Algorithm. 2 Relative to Counterparts}
  \begin{tabular}{cccc}
    \toprule
    \multirow{2}{*}{Scenario} & \multicolumn{3}{c}{Algorithm. 2 vs. Others: Complexity Reduction} \\
    \cmidrule(lr){2-4}
     & (P2.4) & Benchmark & Benchmark w/o $\mathbf{V}$ \\
    \midrule
    \makecell[l]{$M=8,$\\$K^E=K^I=2$}  &  $91.43\%$  &  $94.01\%$  &  $93.34\%$ \\
    \makecell[l]{$M=16,$\\$K^E=K^I=4$} &  $98.54\%$  &  $99.26\%$  &  $99.25\%$ \\
    \addlinespace
    \bottomrule
  \end{tabular}
 \label{table:complexityofalReduction}
\end{table}

\section{Conclusion}\label{sec:V}

In this paper, we proposed a null-space-based scheme for SWIPT and investigated the impact of various waveforms on WET. Our analysis re-assessed the WET strategy, demonstrating that dedicated EBs are not essential, as IBs can concurrently facilitate WET. To support the transmission of deterministic sinusoidal waveforms distinct from the Gaussian waveforms typically employed for WIT, we formulated an optimization problem to derive dedicated EBs and introduced a low-complexity algorithm that omits the WET contribution of IBs, thereby substantially reducing computational complexity. Results reveal that deterministic sinusoidal waveforms outperform Gaussian waveforms in WET efficiency when the input RF power operates within the high-efficiency region of non-linear EH. Numerical results demonstrate that the proposed null-space-based scheme benefits from large-scale antenna systems, laying a solid foundation for SWIPT applications in Massive MIMO scenarios.

{\appendices
\section{Proof of Lemma \ref{lemma:1}}\label{app:A}
Let $r_o=K^I$, we can rewrite Eq. \eqref{eqn:SVDdecompose} as
\begin{align}
	\mathbf{\mathbf{H}}_I
		&=\mathbf{U}_{r_o\times r_o}\left[\begin{array}{ll}
	\boldsymbol{\Sigma}_{1,r_o\times r_o} & \mathbf{0}_{r_o\times \left(M-r_o\right)}
	\end{array}\right]\nonumber\\
	&\qquad\qquad\qquad\times\left[\begin{array}{ll}
	\mathbf{V}_{1,M\times r_o} & \mathbf{V}_{0, M\times\left(M-r_o\right)}
	\end{array}\right]^{\mathrm{H}}\nonumber\\
	&=\mathbf{U}_{r_o\times r_o}
	\boldsymbol{\Sigma}_{1,r_o\times r_o}\mathbf{V}_{1,M\times r_o}^{\mathrm{H}},
\end{align}

when $r^E\geq r_o$, we have
\begin{align}
	&\mathbf{V}_{1,M\times r^E}=\left[\begin{array}{cc}
	 \mathbf{V}_{1,M\times r_o}& \left[\mathbf{V}_{0,M\times\left(M-r_o\right)}\right]_{(:, 1:r^E-r_o)}
	\end{array}\right],\label{eqn:appV1}\\
	&\mathbf{V}_{0,M\times \left(M-r^E\right)}= \left[\mathbf{V}_{0,M\times\left(M-r_o\right)}\right]_{(:, r^E-r_o+1:M-r_o)},\label{eqn:appV0}
\end{align}
where $[\mathbf{A}]_{(:, a:b)}$ denotes the $a$-th column to $b$-th column of matrix $\mathbf{A}$. Eq. \eqref{eqn:appV1} and Eq. \eqref{eqn:appV0} illustrates that $r^E\geq r_o$, $\mathbf{V}_{1,M\times r^E}$ contains not only all the column vectors of $\mathbf{V}_{1,M\times r_o}$ but also the first $r^E-r_o$ columns of $\mathbf{V}_{0, M\times\left(M-r_o\right)}$. Thus, there is no intersection between the column vectors of $\mathbf{V}_{1,M\times r_o}$ and $\mathbf{V}_{0,M\times \left(M-r^E\right)}$, resulting in
\begin{align}
	&\mathbf{H}_I\mathbf{V}_{0,M\times \left(M-r^E\right)}\nonumber\\
	&=\mathbf{U}_{r_o\times r_o}
\boldsymbol{\Sigma}_{1,r_o\times r_o}\mathbf{V}_{1,M\times r_o}^{\mathrm{H}}\mathbf{V}_{0,M\times \left(M-r^E\right)}\nonumber\\
&=\mathbf{0}_{r_o\times \left(M-r^E\right)}. \label{eqn:appInterelimini}
\end{align}
Eq. \eqref{eqn:appInterelimini} demonstrates that the interference signals will be eliminated when $r^E\geq r_o$. As for $r^E < r_o$, we have
\begin{align}
	&\mathbf{H}_I\mathbf{V}_{0,M\times\left(M-r^E\right)}\nonumber\\
	&=\mathbf{A}\left[
	\begin{array}{cc}
	\left[\mathbf{V}_{1,M\times r_o}\right]_{\left(:,r^E+1:r_o\right)}
	&\mathbf{V}_{0, M\times\left(M-r_o\right)}
	\end{array}
	\right]\nonumber\\
	&=\mathbf{U}_{r_o\times r_o}
	\boldsymbol{\Sigma}_{1,r_o\times r_o}\mathbf{V}_{1,M\times r_o}^{\mathrm{H}}\nonumber\\
	&\qquad\qquad\qquad\times\left[
	\begin{array}{cc}
	\left[\mathbf{V}_{1,M\times r_o}\right]_{\left(:,r^E+1:r_o\right)}
	&\mathbf{V}_{0, M\times\left(M-r_o\right)}
	\end{array}
	\right]\nonumber\\
	&=\left[
	\begin{array}{cc}
	\mathbf{U}_{r_o\times r_o}
	\boldsymbol{\Sigma}_{1,r_o\times r_o}\left[
	\begin{array}{c}
	\mathbf{0}_{r^E\times \left(r_o-r^E\right)}\nonumber\\
	\mathbf{I}_{\left(r_o-r^E\right)\times \left(r_o-r^E\right)}
	\end{array}
	\right]
	& \mathbf{0}_{r_o\times\left(M-r_o\right)}
	\end{array}
	\right]\nonumber\\
	&=\mathbf{U}_{r_o\times r_o}\left[
	\begin{array}{cc}
	\begin{array}{c}
	\mathbf{0}_{r\times \left(r_o-r^E\right)}\\
	\left[\boldsymbol{\Sigma}_{1,r_o\times r_o}\right]_{\left(r^E+1:r_o,r+1:r_o\right)}
	\end{array}
	& \mathbf{0}_{r_o\times\left(M-r_o\right)}
	\end{array}
	\right]\nonumber\\
	&=\left[
	\begin{array}{cc}
	\left[\mathbf{U}_{r_o\times r_o}\boldsymbol{\Sigma}_{1,r_o\times r_o}\right]_{\left(1:r_o,r^E+1:r_o\right)}
	& \mathbf{0}_{r_o\times\left(M-r_o\right)}
	\end{array}
	\right]\nonumber\\
	&\neq\mathbf{0}, \label{eqn:appNoInterelimini}
\end{align}
which shows that the interference signal has not been completely eliminated. Moreover, an excessive $r^E$ can result in a decrease in energy performance (the relevant proof is provided in Appendix \ref{app:B}, which can be easily proven using the Cauchy interlacing theorem\cite{bellman1997introduction}). In summary, we find that the optimal value is $r^E=r_o=K^I$, and a similar conclusion applies for $r^I=K^I-1$, thereby completing the proof.

\section{Proof of Lemma \ref{lemma:2}}\label{app:B}
	Observe from $\text{(P2.2)}$ that the this problem satisfies Slater's condition, resulting in a zero duality gap. Consequently, we consider the Lagrangian of $\text{(P2.2)}$ expressed as
	\begin{align}
		&\mathcal{L}\left\{\mathbf{S}_{E,i},\mathbf{S}_E,\lambda_i,\beta\right\}=\beta P_{\text{max}}-\sum_{i\in\boldsymbol{\mathcal{K}}^I}\lambda_i\sigma^2_0\nonumber\\
		&\qquad\qquad\qquad\quad+\sum_{i\in\boldsymbol{\mathcal{K}}^I}\text{trace}\left\{\boldsymbol{\mathcal{A}}_i\mathbf{B}_i\right\}+\text{trace}\left\{\boldsymbol{\mathcal{C}}\mathbf{D}\right\},\label{eqn:lag_function}
	\end{align}
	where $\lambda_i$ and $\beta$ are the dual variables associated with the $i$-th SINR constraint and the transmit sum-power constraint. And $\boldsymbol{\mathcal{A}}_i$ and $\boldsymbol{\mathcal{C}}$ can be expressed as
	\begin{align}
		&\boldsymbol{\mathcal{A}}_i=\mathbf{S}_{E,i}+\frac{\lambda_i\mathbf{h}_{I,I}^k\left(\mathbf{h}_{I,I}^k\right)^H}{2^{C_\text{thre}}-1}-\beta\mathbf{I}_{\left(M-r^I\right)\times \left(M-r^I\right)},\forall i\in\boldsymbol{\mathcal{K}}^I,\\
		&\boldsymbol{\mathcal{C}}=\mathbf{S}_E-\beta\mathbf{I}_{\left(M-r^E\right)\times \left(M-r^E\right)},
	\end{align}
	Thus, the dual problem of the original problem can be formulated as:
	\begin{align}
		\text{(D2.2):  }&\min_{\lambda_i\geq0,\beta\geq0}\beta P_{\text{max}}-\sum_{i\in\boldsymbol{\mathcal{K}}^I}\lambda_i\sigma_0^2\label{eqn:D22}\\
		\text{s.t. } \
		&\boldsymbol{\mathcal{C}}\preceq \mathbf{0},\boldsymbol{\mathcal{A}}_i\preceq\mathbf{0},\forall i\in\boldsymbol{\mathcal{K}}^I.\tag{\ref{eqn:D22}a}\label{eqn:conD22}
	\end{align}
	Denote by $\lambda_i^\S$ the optimal solution, we need to consider the following two cases for the solution of $\text{(D2.2)}$:
	\begin{enumerate*}
        \item $\lambda_i^\S=0,\forall i\in\boldsymbol{\mathcal{K}}^I$,
        \item at least one $\lambda_i^\S>0, \forall i\in\boldsymbol{\mathcal{K}}^I$.
    \end{enumerate*}
	
	As for $\lambda_i^\S=0,\forall i\in\boldsymbol{\mathcal{K}}^I$, we have the optimal $\boldsymbol{\mathcal{A}}_i^\S, \forall i\in\boldsymbol{\mathcal{K}}^I$ and $\boldsymbol{\mathcal{C}}^\S$ as
	\begin{align}
		&\boldsymbol{\mathcal{A}}_i^\S=\mathbf{S}_{E, i}-\beta \mathbf{I}_{\left(M-r^I\right)\times \left(M-r^I\right)}, \forall i\in\boldsymbol{\mathcal{K}}^I,\label{eqn:appOptA}\\
		&\boldsymbol{\mathcal{C}}^\S=\mathbf{S}_E-\beta \mathbf{I}_{\left(M-r^E\right)\times \left(M-r^E\right)}.\label{eqn:appOptC}
	\end{align}
	Denote by $\xi_{E,i}^{\max}$ the maximal eigenvalue of matrix $\mathbf{S}_{E,i}$, the maximal eigenvalue (in fact, any eigenvalues of $\mathbf{S}_{E,i}$ is satisfied) of $\left(\mathbf{S}_{E,i}-\beta \mathbf{I}\right)$ is $\xi_{E,i}^{\max}-\beta$, since $\left(\mathbf{S}_{E,i}-\beta \mathbf{I}\right)\mathbf{x}=\mathbf{S}_{E,i}\mathbf{x}-\beta\mathbf{I}\mathbf{x}=\mathbf{S}_{E,i}\mathbf{x}-\beta \mathbf{x}=\xi_{E,i}^{\max}\mathbf{x}-\beta \mathbf{x}$, where $\mathbf{x}$ denotes the eigenvector corresponding to $\xi_{E,i}^{\max}$.

	Since $\boldsymbol{\mathcal{A}}_i^\S\preceq\mathbf{0}$, we have $\xi_{E,i}^{\max}-\beta\leq0$, and similarly we have $\beta\geq\xi_{E}^{\max}$, where $\xi_{E}^{\max}$ denotes the maximal eigenvalue of $\boldsymbol{\mathcal{C}}^\S$. To summarize, we have $\beta\geq\max\left\{\xi^{\max}_{E,n},\xi_E^{\max}\right\}$. This solution, also studied in \cite{JieXu_NoEB} as OeBF-feasible case, cannot occur here (see Appendix A of \cite{JieXu_NoEB}).

	As for at least one $\lambda_i^\S>0, \forall i\in\boldsymbol{\mathcal{K}}^I$, the optimal $\boldsymbol{\mathcal{C}}^\S$ can also be shown in Eq. \eqref{eqn:appOptC}, and
	\begin{align}
		\boldsymbol{\mathcal{A}}^\S_i=\mathbf{S}_{E,i}+\frac{\lambda_i^\S\mathbf{h}_{I,I}^k\left(\mathbf{h}_{I,I}^k\right)^H}{2^{C_\text{thre}}-1}-\beta\mathbf{I}_{\left(M-r^I\right)\times \left(M-r^I\right)}.
	\end{align}

	According to the constraints \eqref{eqn:conD22}, we have $\beta\geq \max \left\{\xi(\boldsymbol{\mathcal{A}}^\S_i), \xi_E^{\max}\right\}$, where $\xi(\mathbf{M})$ denote the dominant eigenvalue of $\mathbf{M}$. We first compare the value of $\xi^{\max}_{E,i}$ and $\xi_E^{\max}$, and rewrite the $\mathbf{S}_{E,i}$ and $\mathbf{S}_E$ as
	\begin{align}
		\mathbf{S}_{E,i}&=\sum_{j\in\boldsymbol{\mathcal{K}}^E}\mathbf{h}_{E,I}^{j,i}\left(\mathbf{h}_{E,I}^{j,i}\right)^H=\sum_{j\in\boldsymbol{\mathcal{K}}^E}\left(\mathbf{N}_{i}^{I}\right)^H\mathbf{h}_{E}^{j}\left(\mathbf{h}_{E}^{j}\right)^H\mathbf{N}_{i}^{I}\nonumber\\
		&=\left(\mathbf{N}_{i}^{I}\right)^H\mathbf{G}\mathbf{N}_{i}^{I},\label{eqn:appReSei}
	\end{align}
	where $\mathbf{G}=\sum_{j\in\boldsymbol{\mathcal{K}}^E}\mathbf{h}_{E}^{j}\left(\mathbf{h}_{E}^{j}\right)^H$. Similarly we have
	\begin{align}
		\mathbf{S}_E=\left(\mathbf{N}^{E}\right)^H\mathbf{G}\mathbf{N}^{E},\label{eqn:appReSe}
	\end{align}
	Observe from Eq. \eqref{eqn:nullNI}-\eqref{eqn:nullNE} and Eq. \eqref{eqn:appReSei}-\eqref{eqn:appReSe} that, for any column vector $\mathbf{v}^H\in\text{null}\left(\mathbf{H}_I^{\boldsymbol{\mathcal{K}}^I}\right)$, we have $\mathbf{v}\left[\mathbf{H}_I^{\boldsymbol{\mathcal{K}}^I}\right]_{:,j}=0,\forall j\in\boldsymbol{\mathcal{K}}^I$ and $\mathbf{v}\left[\mathbf{H}_I^{\boldsymbol{\mathcal{K}}^I\backslash \mathcal{K}_I^i}\right]_{:,j}=0,\forall j\neq i$. Thus, without loss of generality, we can construct the $\mathbf{N}^{I}$ and $\mathbf{N}^{E}$ as
	\begin{align}
		\mathbf{N}^E_{i}=\mathbf{N}^I_{i}\mathbf{P},
	\end{align}
	where $\mathbf{P}=\left[\mathbf{I}_{M-r^E\times M-r^E};\mathbf{0}_{r^E-r^I\times M-r^E}\right]$. And we can build the relationship between $\mathbf{S}_{E,i}$ and $\mathbf{S}_E$ as

	\begin{align}
		\mathbf{S}_{E} &= \left(\mathbf{N}^{E}\right)^H\mathbf{G}\mathbf{N}^{E}\nonumber\\
		&=\mathbf{P}^H\left(\mathbf{N}^I_{i}\right)^H\mathbf{G}\mathbf{N}^I_{i}\mathbf{P}\nonumber\\
		&=\mathbf{P}^H\mathbf{S}_{E,i}\mathbf{P},
		\label{eqn:relationshipSS}
	\end{align}
	Base on Eq. \eqref{eqn:relationshipSS}, according to Cauchy interlacing theorem \cite{bellman1997introduction}, let $\xi^1_{E}\leq\cdots\leq\xi_{E}^{M-r^E}$ are the eigenvalues of $\mathbf{S}_E$ and $\xi^1_{E,i}\leq\cdots\leq\xi_{E,i}^{M-r^I}$ are the eigenvalues of $\mathbf{S}_{E,i}$, we have
	\begin{align}
		\xi_{E,i}^j\leq\xi_{E}^j\leq\xi_{E,i}^{r^E-r^I+j},
	\end{align}
	when $j=M-r^E$, we finally obtain
	\begin{align}
		\xi_{E,i}^{M-r^E}\leq\xi_{E}^{M-r^E}=\xi^{\max}_E\leq\xi_{E,i}^{M-r^I}=\xi^{\max}_{E,i}. \label{eqn:eigvalueEEi}
	\end{align}

	According to Weyl's inequality \cite{horn2012matrix}, we have $\xi_{E,i}^{\max}+ \xi\{\frac{\lambda_i^\S\mathbf{h}_{I,I}^k\left(\mathbf{h}_{I,I}^k\right)^H}{2^{C_\text{thre}}-1}\}\geq \xi_{E,i}^{\max}\geq \xi^{\max}_E$ since $\mathbf{h}_{I,I}^k\left(\mathbf{h}_{I,I}^k\right)^H$ is non-negative definite. And it is easy to derive that $\beta>\xi_E^{\max}$ base on Eq. \eqref{eqn:eigvalueEEi} and $\mathbf{P}=\left[\mathbf{I}_{M-r^E\times M-r^E};\mathbf{0}_{M-r^E\times r^E-r^I}\right]$. Thus, we have $\text{rank}\left(\boldsymbol{\mathcal{C}}^\S\right)=M-r^E$ since $\boldsymbol{\mathcal{C}}^\S=\mathbf{S}_E-\beta\mathbf{I}_{\left(M-r^E\right)\times \left(M-r^E\right)}\prec\mathbf{0}$, resulting in $\mathbf{D}=\mathbf{0}$ as $\boldsymbol{\mathcal{C}}^\S$ span the entire space. Moreover, it was shown in Remark 3.1 of \cite{JieXu_NoEB} that there always exists an optimal solution with $\text{rank}\left(\mathbf{B}_i\right)=1,\forall i\in\boldsymbol{\mathcal{K}}^I$. Thus, the proof is completed.
\section{Proof of Lemma \ref{lemma:3}}\label{app:C}
	Similar with Lemma \ref{lemma:2}, we also consider the Lagrangian of $\text{(P2.4)}$ as Eq. \eqref{eqn:lag_function}, where
	\begin{align}
		&\boldsymbol{\mathcal{A}}^\S_i=\mathbf{S}_{E,i}+\frac{\lambda_i^\S\mathbf{h}_{I,I}^k\left(\mathbf{h}_{I,I}^k\right)^H}{2^{C_\text{thre}}-1}-\beta\mathbf{I}_{\left(M-r^I\right)\times \left(M-r^I\right)},\\
		&\boldsymbol{\mathcal{C}}^\S=\eta\mathbf{S}_E-\beta \mathbf{I}_{\left(M-r^E\right)\times \left(M-r^E\right)},\label{eqn:newC}
	\end{align}
	when at least one $\lambda_i^\S>0, \forall i\in \boldsymbol{\mathcal{K}}^I$, where Eq. \ref{eqn:newC} comes from $\eta\text{trace}\left\{\mathbf{S}_E\mathbf{D}\right\}=\text{trace}\left\{\eta\mathbf{S}_E\mathbf{D}\right\}$.

	According to Lemma \ref{lemma:2} and Appendix \ref{app:B}, we have $\mathbf{D} \neq \mathbf{0}$ when $\xi^{\max}_E\geq \xi^{\max}_{E,i},\forall i\in \boldsymbol{\mathcal{K}}^I$. Thus, we further expressed as
	\begin{align}
		\xi\left\{\eta\mathbf{S}_E\right\}=\eta\xi\left\{\mathbf{S}_E\right\}\geq \xi\left\{\mathbf{S}_{E,i}+\frac{\lambda_i^\S\mathbf{h}_{I,I}^k\left(\mathbf{h}_{I,I}^k\right)^H}{2^{C_\text{thre}}-1}\right\}.
	\end{align}
	
	And since $\xi_{E,i}^{\max}+ \xi\{\frac{\lambda_i^\S\mathbf{h}_{I,I}^k\left(\mathbf{h}_{I,I}^k\right)^H}{2^{C_\text{thre}}-1}\}\geq \xi\left\{\mathbf{S}_{E,i}+\frac{\lambda_i^\S\mathbf{h}_{I,I}^k\left(\mathbf{h}_{I,I}^k\right)^H}{2^{C_\text{thre}}-1}\right\}$ via Weyl's inequality, the bound of $\eta$ can be further expressed as
	\begin{align}
		\eta&\geq \frac{\xi_{E,i}^{\max}+ \xi\{\frac{\lambda_i^\S\mathbf{h}_{I,I}^k\left(\mathbf{h}_{I,I}^k\right)^H}{2^{C_\text{thre}}-1}\}}{\xi\left\{\mathbf{S}_E\right\}}\nonumber\\
		&=\frac{\xi_{E,i}^{\max}+ \xi\{\frac{\lambda_i^\S\mathbf{h}_{I,I}^k\left(\mathbf{h}_{I,I}^k\right)^H}{2^{C_\text{thre}}-1}\}}{\xi_E^{\max}}
		=\frac{\xi_{E,i}^{\max}}{\xi_E^{\max}}+\delta, \forall i\in \boldsymbol{\mathcal{K}}^I,\label{eqn:etavalue}
	\end{align}
	where $\delta = \frac{\xi\{\frac{\lambda_i^\S\mathbf{h}_{I,I}^k\left(\mathbf{h}_{I,I}^k\right)^H}{2^{C_\text{thre}}-1}\}}{\xi_E^{\max}}$. and Eq. \eqref{eqn:etavalue} can be expressed as $\eta > \frac{\max_{i\in\boldsymbol{\mathcal{K}}^I}(\xi^{\max}_{E,i})}{\xi^{\max}_E}+\delta$ for any available $i\in\boldsymbol{\mathcal{K}}^I$. In general, $\delta$ is a small value due to the denominator component $\left(2^{C_\text{thre}}-1\right)\xi_E^{\max}$. Thus, the proof is completed.
}
\bibliography{Reference}
\end{document}